\documentclass[11pt,preprint]{aastex}

\usepackage{amsmath,amssymb,amscd}
\usepackage{graphicx}
\usepackage{txfonts}
\usepackage{natbib}
\usepackage{braket}

\shorttitle{Modeling the scattering polarization of the Na~{\sc i} D$_1$ line}
\shortauthors{Belluzzi, Trujillo Bueno \& Landi Degl'Innocenti}

\begin{document}

\title{Radiative transfer modeling of the enigmatic\\
scattering polarization in the solar Na~{\sc i} D$_1$ line}

\author{{\sc Luca Belluzzi}\altaffilmark{1,2},
{\sc Javier Trujillo Bueno}\altaffilmark{3,4,5},
{\sc and Egidio Landi Degl'Innocenti}\altaffilmark{6}}
\altaffiltext{1}{Istituto Ricerche Solari Locarno, CH-6605 Locarno Monti, 
Switzerland}
\altaffiltext{2}{Kiepenheuer-Institut f\"ur Sonnenphysik, D-79104 Freiburg, 
Germany}
\altaffiltext{3}{Instituto de Astrof\'isica de Canarias, E-38205 La Laguna, 
Tenerife, Spain}
\altaffiltext{4}{Departamento de Astrof\'isica, Facultad de F\'isica, 
Universidad de La Laguna, E-38206 La Laguna, Tenerife, Spain}
\altaffiltext{5}{Consejo Superior de Investigaciones Cient\'ificas, Spain}
\altaffiltext{6}{Dipartimento di Fisica e Astronomia, Universit\`a di Firenze, 
I-50125 Firenze, Italy}

\begin{abstract}
The modeling of the peculiar scattering polarization signals observed in some 
diagnostically important solar resonance lines requires the consideration of 
the detailed spectral structure of the incident radiation field as well as the 
possibility of ground level polarization, along with the atom's hyperfine 
structure and quantum interference between hyperfine $F$-levels pertaining 
either to the same fine structure $J$-level, or to different $J$-levels of the 
same term.
Here we present a theoretical and numerical approach suitable for solving this 
complex non-LTE radiative transfer problem.
This approach is based on the density-matrix metalevel theory (where each level 
is viewed as a continuous distribution of sublevels) and on accurate formal 
solvers of the transfer equations and efficient iterative methods. 
We show an application to the D-lines of Na~{\sc i}, with emphasis on the 
enigmatic D$_1$ line, pointing out the observable signatures of the various 
physical mechanisms considered. 
We demonstrate that the linear polarization observed in the core of the 
D$_1$ line may be explained by the effect that one gets when the detailed 
spectral structure of the anisotropic radiation responsible for the optical 
pumping is taken into account.
This physical ingredient is capable of introducing significant scattering 
polarization in the core of the Na~{\sc i} D$_1$ line without the need for 
ground-level polarization.
\end{abstract}


\section{Introduction}
For nearly two decades, the modeling of the scattering polarization signals 
observed by \citet{Ste97} in the core of the Na~{\sc i} and Ba~{\sc ii} D$_1$ 
lines has represented one of the most challenging problems in the field of 
theoretical spectropolarimetry \citep[see also][]{Ste00}. 
Given that these spectral lines are produced by atomic transitions between an 
upper level and a lower level with total angular momentum $J=1/2$ (i.e., atomic 
levels that cannot carry atomic alignment), they were initially considered to
be intrinsically unpolarizable lines.
On the other hand, the only stable isotope of sodium ($^{23}$Na) and two 
of the seven stable isotopes of barium ($^{135}$Ba and $^{137}$Ba, with 
relative abundances of about 7\% and 11\%, respectively) have nuclear spin 
$I=3/2$.
In these isotopes, the upper and lower levels of the D$_1$ line split into 
two hyperfine structure (HFS) levels with total (electronic plus nuclear) 
angular momenta $F=1$ and $F=2$.
Therefore, the main problem is identifying a physical mechanism through 
which atomic alignment can be induced in the $F$-levels of the D$_1$ line.

According to the theory of spectral line polarization described in 
\citet{Lan04}, the mere absorption of anisotropic radiation is not sufficient 
to induce atomic alignment in the upper $F$-levels of the D$_1$ line, unless 
the lower $F$-levels are also polarized.
As shown in Sect.~2 of \citet{Bel13b}, this is ultimately due to the 
hypothesis, required by the theory of \citet{Lan04}, that the incident 
radiation field is flat (i.e., independent of frequency) across the spectral 
interval spanned by the HFS components of the D$_1$ line (flat-spectrum 
approximation).
This assumption, on the other hand, does not appear to be particularly 
unsuitable in the solar case, since the frequency separation among the various 
HFS components of the D$_1$ line is significantly smaller than the Doppler 
width of the spectral line.

By generalizing the idea of ``internal levels'' (or ``metalevels'') to the 
polarized case, \citet{Lan97} developed a theoretical approach that does not 
require the flat-spectrum condition to be satisfied, and that is suitable for 
treating coherent scattering processes in the presence of pumping radiation 
fields with an arbitrary spectral structure. 
Working within the framework of this theory, and assuming that the anisotropy 
degree of the incident radiation is constant with frequency, 
\citet{Lan98,Lan99} showed that a conspicuous polarization signal, similar to 
the observed one, can be produced in the core of the Na~{\sc i} D$_1$ line, 
provided that a substantial amount of atomic polarization is present in the 
lower level (the ground level of sodium).
This result, on the other hand, leads to a sort of paradox since the required 
atomic polarization in the long-lived ground level of sodium is incompatible 
with the presence in the lower solar chromosphere of inclined magnetic fields 
sensibly stronger than 0.01~G \citep[see][]{Lan98}, which seems to contradict 
the results obtained from other types of observations 
\citep[e.g.,][]{Bia98,Ste98}.
Moreover, through a calculation based on Quantum Chemistry, \citet{Ker02} 
argued that the effect of depolarizing collisions is sufficiently strong to 
destroy the required atomic polarization in the ground level of sodium.

In the work of \citet{Lan98}, lower level polarization was included in the 
problem as a free parameter.
The physical mechanism through which atomic polarization can be induced in the 
ground level of sodium via the D$_2$ line transition, and then transferred 
to the upper $F$-levels of the D$_1$ line (the repopulation pumping mechanism), 
was pointed out by \citet{JTB02}, who in addition investigated the sensitivity 
of the atomic polarization of the sodium HFS levels to the presence of magnetic 
fields \citep[see also][]{Cas02}.
These works, which were carried out within the framework of the theory of 
polarization described in \citet{Lan04}, neglecting depolarizing collisions, 
showed the actual possibility of inducing a significant amount of atomic 
polarization in the ground level of sodium, but also confirmed its 
incompatibility with the estimated intensities and geometries of the magnetic 
fields of the lower chromosphere.

The ``enigma'' of the D$_1$ lines remained substantially unchanged until the 
recent identification by \citet{Bel13b} and by \citet{DPA14} of two mechanisms 
that can introduce scattering polarization in the core of such lines,
without requiring the presence of atomic polarization in the ground levels of 
sodium and barium.
The idea at the basis of the mechanism identified by \citet{Bel13b} is that if 
the incident radiation field (and, in particular, its anisotropy degree) 
varies across the HFS multiplet, so that the various HFS components are 
affected by different pumping radiations, then atomic polarization can be 
induced in the upper $F$-levels of the D$_1$ line, and the emitted radiation 
can in general be polarized, also in the absence of atomic polarization in the 
lower $F$-levels.
\citet{Bel13b} modeled the D$_1$ lines of Na~{\sc i} and Ba~{\sc ii} by solving 
the full non-LTE radiative transfer (RT) problem in one-dimensional 
semi-empirical models of the solar atmosphere, according to the partial 
frequency redistribution (PRD) approach described in \citet{Bel14}. 
Through the $R_{II}$ part of their redistribution matrix, which describes 
coherent scattering processes according to the metalevel approach, they
took into account the detailed spectral structure of the pumping radiation, 
finding that the small differences among the radiation fields experienced by 
the various HFS components of the D$_1$ line are actually sufficient to produce 
appreciable scattering polarization signals in the core of these lines.

The signal obtained by \citet{Bel13b} in the core of the Na~{\sc i} D$_1$ line 
is, however, sensibly weaker than the one observed by \citet{Ste97}, but it is 
similar to that shown in the right panel of Figure~2 of \citet{JTB09}, 
resulting from the observations by \citet{JTB01}.
On the other hand, two physical ingredients, which were taken into account by 
\citet{Lan98}, have been neglected by \citet{Bel13b}: quantum interference 
between HFS magnetic sublevels pertaining to different $J$-levels, and the 
possibility that a given amount of atomic polarization is present in the lower 
$F$-levels.
It is well known that interference between the upper $J$-levels of the D$_1$ 
and D$_2$ lines plays a very important role in the generation of the scattering 
polarization pattern observed in these lines \citep[see][]{Ste80,Lan98,Lan04},
while recent RT calculations seem to indicate that in the atmospheric region 
where the core of the sodium D-lines is formed depolarizing collisions might 
not completely destroy lower level polarization as previously thought.

In the first part of this paper, we derive a redistribution matrix suitable for 
describing coherent scattering in the atom rest frame (with Doppler 
redistribution in the observer's frame), accounting for lower level 
polarization, interference between HFS magnetic sublevels pertaining 
to the upper $J$-levels of D$_1$ and D$_2$, and inelastic collisions with 
electrons. 
In the second part, we describe the numerical method of solution of the ensuing 
non-LTE problem, and we present a series of results obtained by treating
lower level polarization as a free parameter of the problem (it will not 
be calculated self-consistently when solving the non-LTE problem). 

As previously mentioned, another mechanism that may explain the physical 
origin of the signals observed by \citet{Ste97} and \citet{Ste00} in the core 
of the Na~{\sc i} and Ba~{\sc ii} D$_1$ lines, without requiring the presence 
of lower level polarization, has been recently identified by \citet{DPA14}.
In their work, the authors show that measurable $Q/I$ signals can be produced 
in the core of intrinsically unpolarizable lines through the redistribution of 
the spectral line radiation due to the non-coherence of the continuum 
scattering.
Although this mechanism (which strongly depends on the assumed model 
atmosphere) may well coexist with the previous one, it will be neglected in 
this investigation.

Finally, we point out that the enigmatic polarization signals observed in the 
solar D$_1$ line have led to the realization of a laboratory experiment on 
scattering polarization by the potassium atom, which has the same D$_1$ quantum 
structure as sodium \citep{Tha06,Tha09}.
This experiment, performed by pumping the potassium atoms through a tunable 
laser, has shown an unexpected phenomenology that cannot be interpreted by 
means of the standard Kramers-Heisenberg equation.
\citet{Ste15} has recently suggested an interpretation of these results based 
on quantum interference between the sublevels of the ground state of potassium.
It will be of interest to investigate whether this physical ingredient may also 
play a role in the solar case, but this lies outside the scope of the present 
paper. This kind of interference is thus neglected in the present work.

\section{Formulation of the problem}
We consider a two-term atom with HFS, in the absence of magnetic fields.
Under the assumption of $L$-$S$ coupling, the atomic Hamiltonian has 
eigenvectors of the form $\ket{ \, \beta L S I J F M \,}$, where $\beta$, $L$, 
and $S$ indicate the electronic configuration, the orbital angular momentum, 
and the electronic spin, respectively, $I$ is the nuclear spin, $J$ the total 
electronic angular momentum, $F$ the total (electronic plus nuclear) angular 
momentum, and $M$ its projection along the quantization axis.
We recall that each term is composed of ($L+S-|L-S|+1$) fine structure (FS) 
$J$-levels, while each $J$-level splits into ($J+I-|J-I|+1$) HFS $F$-levels.
In each term, we thus have
\begin{equation}
	\sum^{L+S}_{J=|L-S|} \left( \sum^{J+I}_{F=|J-I|} (2F + 1) \right) = 
	(2L + 1)(2S + 1)(2I + 1) \; ,
\end{equation}
magnetic sublevels.

In the standard representation, the atomic model under consideration is 
described by the density matrix elements
\begin{equation}
	^{\beta L S I}\rho(J F M, J^{\prime} F^{\prime} M^{\prime}) \equiv
	\bra{ \, \beta L S I J F M } \, \hat{\rho} \, 
	\ket{ \, \beta L S I J^{\prime} F^{\prime} M^{\prime}} \; ,
\label{Eq:density_matrix}
\end{equation}
with $\hat{\rho}$ the density operator.
As is clear from Eq.~(\ref{Eq:density_matrix}), this atomic model accounts for 
quantum interference between pairs of HFS magnetic sublevels belonging either 
to the same $F$-level or to different $F$-levels pertaining either to the same 
$J$-level or to different $J$-levels within the same term.
In the following, we will work with the irreducible spherical components of the 
density matrix (or spherical statistical tensors) defined by \citep[see 
Eq.~(11b) of][]{Cas05}
\begin{equation}
	{^{\beta L S I}\rho^K_Q}(J F, J^{\prime} F^{\prime}) =
	\sum_{M M^{\prime}} (-1)^{F-M} \sqrt{2K + 1} 
	\left( \! \begin{array}{ccc}
		F & F^{\prime} & K \\
		M & -M^{\prime} & -Q 
	\end{array} \! \right) \,
	{^{\beta L S I}\rho}(J F M, J^{\prime} F^{\prime} M^{\prime}) \; .
\label{Eq:spher_stat_tens}
\end{equation}
Equation~(\ref{Eq:spher_stat_tens}) shows that in a two-term (or multi-term) 
atom with HFS, besides the 0-rank elements $^{\beta L S I}\rho^0_0(J F, J F)$, 
which are proportional to the population of a given HFS level, we also have 
non-diagonal 0-rank elements of the form $^{\beta L S I}\rho^0_0(J F, 
J^{\prime} F)$, which describe interference between pairs of magnetic sublevels 
with the same $M$ and $F$ quantum numbers, but pertaining to different 
$J$-levels.

We take lower term polarization into account, but only in the form of 
population imbalances among the various magnetic sublevels of the same 
$F$-level (interference between pairs of magnetic sublevels pertaining either 
to the same $F$-level or to different $F$-levels of the lower term is instead 
neglected by definition).\footnote{Note that in the recent modeling of 
polarized scattering by tunable laser light on potassium gas in the laboratory, 
quantum coherence in the lower term was identified as the source of the 
observed D$_1$ polarization \citep{Ste15}.
It will be of interest to investigate whether this coherence may play a 
significant role also in the solar case, where the pumping radiation is 
broadband and not nearly monochromatic and highly polarized (as in a tunable 
laser).
In this work, we demonstrate that it is possible to obtain significant solar 
D$_1$ polarization by neglecting them.}
Within the framework of the density matrix formalism, this hypothesis reads
\begin{equation}
	{^{\beta_{\ell} L_{\ell} S I}\rho^K_Q}(J_{\ell} F_{\ell}, 
	J^{\prime}_{\ell} F^{\prime}_{\ell}) = 
	\delta_{J_{\ell} J^{\prime}_{\ell}} \, 
	\delta_{F_{\ell} F^{\prime}_{\ell}} \, \delta_{Q 0} \; 
	{^{\beta_{\ell} L_{\ell} S I}\rho^K_0}(J_{\ell} F_{\ell}, J_{\ell} 
	F_{\ell}) \; ,
\label{Eq:rho_lt_1}
\end{equation}
where the label $\ell$ indicates that the corresponding quantity refers to the 
lower term, or to one of its FS or HFS levels (accordingly, quantities of the 
upper term will be labeled with the letter $u$).
Furthermore, we assume that the populations of the various HFS $F$-levels of 
the lower term, $\mathcal{N}(J_{\ell}, F_{\ell})$, are proportional to the 
statistical weights:
\begin{equation}
	\mathcal{N}(J_{\ell}, F_{\ell}) =
	\frac{\mathcal{N}(L_{\ell})}{(2L_{\ell} + 1)(2S + 1)(2I + 1)}
	(2F_{\ell} + 1) \; ,
\end{equation}
with $\mathcal{N}(L_{\ell})$ the overall population of the lower term.
Indicating with $\mathcal{N}_T$ the total population, we have
\begin{equation}
	{^{\beta_{\ell} L_{\ell} S I}\rho^0_0}(J_{\ell} F_{\ell}, J_{\ell} 
	F_{\ell}) = \frac{1}{\mathcal{N}_T} 
	\frac{\mathcal{N}(J_{\ell}, F_{\ell})}{\sqrt{2 F_{\ell} + 1}} 
	= \frac{\mathcal{N}(L_{\ell})}{\mathcal{N}_T}
	\frac{\sqrt{2F_{\ell} + 1}}{(2L_{\ell} + 1)(2S + 1)(2I + 1)} \;  .
\label{Eq:rho00}
\end{equation}
Introducing the quantity
\begin{equation}
	\sigma^K_0(J_{\ell}, F_{\ell}) = 
	\frac{ {^{\beta_{\ell} L_{\ell} S I}\rho^K_0}(J_{\ell} F_{\ell}, 
	J_{\ell} F_{\ell}) }
	{ {^{\beta_{\ell} L_{\ell} S I}\rho^0_0}(J_{\ell} F_{\ell}, J_{\ell} 
	F_{\ell}) } \; ,
\label{Eq:sigmaK0}
\end{equation}
we can finally write
\begin{equation}
	{^{\beta_{\ell} L_{\ell} S I}\rho^K_Q}(J_{\ell} F_{\ell}, 
	J^{\prime}_{\ell} F^{\prime}_{\ell}) = 
	\delta_{J_{\ell} J^{\prime}_{\ell}} \, 
	\delta_{F_{\ell} F^{\prime}_{\ell}} \, \delta_{Q 0} \,
	\frac{\mathcal{N}(L_{\ell})}{\mathcal{N}_T}
	\frac{\sqrt{2F_{\ell} + 1}}{(2L_{\ell} + 1)(2S + 1)(2I + 1)} \,
	\sigma^K_0(J_{\ell}, F_{\ell}) \;  .
\label{Eq:rho_lt_2}
\end{equation}
In this work, the quantity $\sigma^K_0(J_{\ell}, F_{\ell})$ is not calculated 
through a self-consistent solution of the statistical equilibrium equations 
and of the RT equations, but it is set as a free parameter of the problem.

We account for inelastic collisions with electrons, inducing transitions 
between the upper and the lower term.
On the other hand, we neglect ``weakly inelastic'' collisions inducing 
transitions between different FS or HFS levels pertaining to the same term, 
as well as elastic collisions with neutral perturbers.
Finally, we assume that the magnetic sublevels of the lower term are infinitely 
sharp. In the atom rest frame, this last set of hypotheses implies coherence in
frequency for Rayleigh scattering, and the usual energy conservation 
relationship between the frequencies of the incoming and outgoing photons 
for Raman scattering.
Stimulated emission is neglected.

\section{The absorption and emission coefficients}
The expression of the absorption coefficient for a two-term atom with HFS, 
accounting for lower term polarization and for the presence of an arbitrary 
magnetic field, is given by Eq.~(33a) of \citet{Cas05}.
With the help of Eq.~(2.34) of \citet{Lan04}, it can be shown that in the 
absence of magnetic fields, it takes the simpler form
\allowdisplaybreaks
\begin{align}
	\eta_i(\nu,\vec{\Omega}) = & \frac{h \nu}{4 \pi} \mathcal{N}_T
	(2L_{\ell} + 1) B(L_{\ell} \rightarrow L_u) 
	\sum_{KQ} \sum_{J_{\ell} F_{\ell}} \sum_{J_{\ell}^{\prime} 
	F_{\ell}^{\prime}} \sum_{J_u F_u} (-1)^{1 + K + J_{\ell}^{\prime} - 
	J_{\ell} - F_{\ell} - F_u} \nonumber \\
	& \times \, (2J_u + 1) (2F_u + 1) 
	\sqrt{3 (2J_{\ell} + 1) (2J_{\ell}^{\prime} + 1) (2F_{\ell} + 1)
	(2F_{\ell}^{\prime} + 1)} \nonumber \\
	& \times \,
	\left\{ \begin{array}{ccc}
		J_u & J_{\ell} & 1 \\
		F_{\ell} & F_u & I 
	\end{array} \right\}
	\left\{ \begin{array}{ccc}
		J_u & J_{\ell}^{\prime} & 1 \\
		F_{\ell}^{\prime} & F_u & I
	\end{array} \right\}
	\left\{ \begin{array}{ccc}
		L_u & L_{\ell} & 1 \\
		J_{\ell} & J_u & S
	\end{array} \right\}
	\left\{ \begin{array}{ccc}
		L_u & L_{\ell} & 1 \\
		J_{\ell}^{\prime} & J_u & S
	\end{array} \right\}
	\left\{ \begin{array}{ccc}
		1 & 1 & K \\
		F_{\ell} & F_{\ell}^{\prime} & F_u
	\end{array} \right\} \nonumber \\
	& \times \,
	\, {\rm Re} \left\{ \mathcal{T}^K_Q(i,\vec{\Omega}) \, 
	{^{\beta_{\ell} L_{\ell} S I}\rho^K_Q}(J_{\ell} F_{\ell}, 
	J_{\ell}^{\prime} F_{\ell}^{\prime}) \, 
	\Phi(\nu_{J_u F_u, J_{\ell} F_{\ell}} - \nu) \right\} \; ,
\label{Eq:eta_1}
\end{align}
with $i=0,1,2,3$, standing for Stokes $I$, $Q$, $U$, and $V$, respectively, and 
where $B(L_{\ell} \rightarrow L_u)$ is the Einstein coefficient for absorption 
from the lower to the upper term and $\mathcal{T}^K_Q(i,\vec{\Omega})$ is the 
geometrical tensor introduced by \citet{Lan83}. 
The complex profile $\Phi(\nu_0 - \nu)$ is defined by
\begin{equation}
	\Phi(\nu_0 - \nu) = \phi(\nu_0 - \nu) +{\rm i} \, \psi(\nu_0 -\nu) \; ,
\label{Eq:Phi_atom}
\end{equation}
with $\phi(\nu_0 - \nu)$ the Lorentzian profile and $\psi(\nu_0 - \nu)$ the 
associated dispersion profile.

Assuming that no interference is present in the lower term, so that the 
spherical statistical tensors of this term are given by 
Eq.~(\ref{Eq:rho_lt_1}), then Eq.~(2) of \citet{Lan99} is easily recovered.
Substituting Eq.~(\ref{Eq:rho_lt_2}) into Eq.~(\ref{Eq:eta_1}), we have
\begin{equation}
	\eta_i(\nu,\vec{\Omega}) = k_L \sum_{K} 
	\mathcal{T}^{K}_0(i,\vec{\Omega}) \, \alpha^K_0(\nu) \; ,
\label{Eq:eta_2}
\end{equation}
with
\allowdisplaybreaks
\begin{align}
	\alpha^K_0(\nu) = & \, \frac{1}{(2S + 1)(2I + 1)} \, 
	\sum_{J_{\ell} F_{\ell}} \sum_{J_u F_u} 
	(-1)^{1 + K - F_{\ell} - F_u} \, \sqrt{3} \, 
	(2J_u + 1)(2J_{\ell} + 1)(2F_u + 1)(2F_{\ell} + 1)^{3/2} \nonumber \\
	& \times \,
	\left\{ \begin{array}{ccc}
		J_u & J_{\ell} & 1 \\
		F_{\ell} & F_u & I 
	\end{array} \right\}^2
	\left\{ \begin{array}{ccc}
		L_u & L_{\ell} & 1 \\
		J_{\ell} & J_u & S
	\end{array} \right\}^2
	\left\{ \begin{array}{ccc}
		1 & 1 & K \\
		F_{\ell} & F_{\ell} & F_u
	\end{array} \right\} \,	\sigma^K_0(J_{\ell}, F_{\ell}) \,
	\phi(\nu_{J_u F_u, J_{\ell} F_{\ell}} - \nu) \; ,
\end{align}
and where we have introduced the frequency-integrated absorption coefficient
\begin{equation}
	k_L = \frac{h \nu}{4 \pi} \, \mathcal{N}(L_{\ell}) \, 
	B(L_{\ell} \rightarrow L_u) \; .
\end{equation}

The expression of the emission coefficient for a two-term atom with HFS, in 
the absence of magnetic fields, neglecting any kind of collisions, assuming 
that no interference is present in the lower term, and under the assumption 
that the magnetic sublevels of the lower term are infinitely sharp, has 
been derived by \citet{Lan99}, working within the framework of the metalevel 
approach of \citet{Lan97}.
This expression, which is given by Eq.~(1) of \citet{Lan99}, can be rewritten 
in the equivalent form
\allowdisplaybreaks
\begin{align}
	\varepsilon_i(\nu, \vec{\Omega}) = & \frac{h \nu}{4 \pi} \, 
	\mathcal{N}_T \, B(L_{\ell} \rightarrow L_u) \,
	(2L_u + 1) (2L_{\ell} + 1) \nonumber \\ 
	& \times \, 
	\sum_{J_{\ell} J^{\prime}_{\ell}} \, \sum_{J_u J^{\prime}_u} \, 
	\sum_{F_{\ell} F^{\prime}_{\ell}} \, \sum_{F_u F^{\prime}_u} \, 
	\sum_{K Q} \, \sum_{K_r Q_r} \, \sum_{K_{\ell}} \, 
	(-1)^{1 + K_{\ell} + F_{\ell} + F_u} \,
	3 \sqrt{(2K + 1)(2K_r + 1)(2K_{\ell} + 1)} \nonumber \\
	& \times \, (2J_u + 1)(2J^{\prime}_u + 1)(2J_{\ell} + 1)
	(2J^{\prime}_{\ell} + 1)(2F_u + 1)(2F^{\prime}_u + 1)(2F_{\ell} + 1)
	(2F^{\prime}_{\ell} + 1) \nonumber \\
	& \times \,
	\left\{ \begin{array}{ccc}
		L_u & L_{\ell} & 1 \\
		J_{\ell} & J_u & S 
	\end{array} \right\}
	\left\{ \begin{array}{ccc}
		L_u & L_{\ell} & 1 \\
		J_{\ell} & J^{\prime}_u & S
	\end{array} \right\}
	\left\{ \begin{array}{ccc}
		L_u & L_{\ell} & 1 \\
		J^{\prime}_{\ell} & J_u & S
	\end{array} \right\}
	\left\{ \begin{array}{ccc}
		L_u & L_{\ell} & 1 \\
		J^{\prime}_{\ell} & J^{\prime}_u & S
	\end{array} \right\} \nonumber \\
	& \times \,
	\left\{ \begin{array}{ccc}
		J_u & J_{\ell} & 1 \\
		F_{\ell} & F_u & I
	\end{array} \right\}
	\left\{ \begin{array}{ccc}
		J^{\prime}_u & J_{\ell} & 1 \\
		F_{\ell} & F^{\prime}_u & I
	\end{array} \right\}
	\left\{ \begin{array}{ccc}
		J_u & J^{\prime}_{\ell} & 1 \\
		F^{\prime}_{\ell} & F_u & I
	\end{array} \right\}
	\left\{ \begin{array}{ccc}
		J^{\prime}_u & J^{\prime}_{\ell} & 1 \\
		F^{\prime}_{\ell} & F^{\prime}_u & I
	\end{array} \right\} \nonumber \\
	& \times \,
	\left\{ \begin{array}{ccc}
		K & F_u & F^{\prime}_u \\
		F_{\ell} & 1 & 1
	\end{array} \right\}
	\left\{ \begin{array}{ccc}
		K & K_r & K_{\ell} \\
		F^{\prime}_u & 1 & F^{\prime}_{\ell} \\
		F_u & 1 & F^{\prime}_{\ell} 
	\end{array} \right\}
	\left( \begin{array}{ccc}
		K & K_r & K_{\ell} \\
		-Q & -Q_r & 0
	\end{array} \right) \nonumber \\
	& \times \,
	\, \mathcal{T}^K_Q(i,\vec{\Omega}) \, J^{K_r}_{Q_r}(\nu - 
	\nu_{J^{\prime}_{\ell} F^{\prime}_{\ell}, J_{\ell} F_{\ell}} ) \, 
	{^{\beta_{\ell} L_{\ell} S I}\rho^{K_{\ell}}_0}(J^{\prime}_{\ell} 
	F^{\prime}_{\ell}, J^{\prime}_{\ell} F^{\prime}_{\ell}) \,
	\frac{1}{2} \left[ \frac{\Phi(\nu_{J_u F_u, J_{\ell} F_{\ell}} - \nu) + 
	\Phi(\nu_{J^{\prime}_u F^{\prime}_u, J_{\ell} F_{\ell}} - \nu)^{\ast}}
	{1 + 2 \pi {\rm i} \nu_{J^{\prime}_u F^{\prime}_u, J_u F_u}/
	A(L_u \rightarrow L_{\ell})} \right] \; ,
\label{Eq:eps_CS_1}
\end{align}
where $A(L_u \rightarrow L_{\ell})$ is the Einstein coefficient for spontaneous 
emission from the upper to the lower term, while $\nu_{ab}$ is the frequency 
separation between levels $a$ and $b$.
The tensor $J^{K_r}_{Q_r}(\nu^{\prime})$ describing the incident radiation 
field is given by \citep[see Eq.~(5.157) of][]{Lan04}
\begin{equation}
	J^{K_r}_{Q_r}(\nu^{\prime}) = \oint \frac{{\rm d} {\Omega}^{'}}{4 \pi} 
	\sum_{j=0}^3 \mathcal{T}^{K_r}_{Q_r}(j,\vec{\Omega}^{\prime}) \, 
	I_j(\nu^{\prime},\vec{\Omega}^{\prime}) \; ,
\label{Eq:JKQ}
\end{equation}
with $I_j(\nu^{\prime},\vec{\Omega}^{\prime})$ the four Stokes parameters.

In order to take the effect of inelastic collisions with electrons into 
account, we proceed by analogy with the complete frequency redistribution (CRD) 
case.
The expression of the emission coefficient for a two-term atom with HFS, 
in the limit of CRD, taking such collisions into account, has been derived by 
\citet{Bel15}, working within the framework of the theory of \citet{Lan04}.
By analogy with Eq.~(37) of \citet{Bel15}, we thus include the effect of 
inelastic collisions by modifying Eq.~(\ref{Eq:eps_CS_1}) as follows
\allowdisplaybreaks
\begin{align}
	\varepsilon_i(\nu, \vec{\Omega}) = & \frac{h \nu}{4 \pi} \, 
	\mathcal{N}_T \, B(L_{\ell} \rightarrow L_u) \,
	(2L_u + 1) (2L_{\ell} + 1) 
	\nonumber \\ 
	& \times \, 
	\sum_{J_{\ell} J^{\prime}_{\ell}} \, \sum_{J_u J^{\prime}_u} \, 
	\sum_{F_{\ell} F^{\prime}_{\ell}} \, \sum_{F_u F^{\prime}_u} \, 
	\Upsilon(L_u L_{\ell} S I, J_u J^{\prime}_u J_{\ell} J^{\prime}_{\ell} 
	F_u F^{\prime}_u F_{\ell} F^{\prime}_{\ell}) \,
	\frac{1}{2} \left[ \frac{\Phi(\nu_{J_u F_u, J_{\ell} F_{\ell}} - \nu) + 
	\Phi(\nu_{J^{\prime}_u F^{\prime}_u, J_{\ell} F_{\ell}} - \nu)^{\ast}}
	{1 + \epsilon^{\prime} + 2 \pi {\rm i} \nu_{J^{\prime}_u F^{\prime}_u, 
	J_u F_u}/A(L_u \rightarrow L_{\ell})} \right] 
	\nonumber \\
	& \times \, 
	\sum_{K Q} \, \sum_{K_r Q_r} \, \sum_{K_{\ell}} \, 
	(-1)^{1 + K_{\ell} + F_{\ell} + F_u} \,
	3 \sqrt{(2K + 1)(2K_r + 1)(2K_{\ell} + 1)} \,
	\nonumber \\
	& \times \,
	\left\{ \begin{array}{ccc}
		K & F_u & F^{\prime}_u \\
		F_{\ell} & 1 & 1
	\end{array} \right\}
	\left\{ \begin{array}{ccc}
		K & K_r & K_{\ell} \\
		F^{\prime}_u & 1 & F^{\prime}_{\ell} \\
		F_u & 1 & F^{\prime}_{\ell} 
	\end{array} \right\}
	\left( \begin{array}{ccc}
		K & K_r & K_{\ell} \\
		-Q & -Q_r & 0
	\end{array} \right) 
	\nonumber \\
	& \times \,
	\, \mathcal{T}^K_Q(i,\vec{\Omega}) \, 
	J^{K_r}_{Q_r}(\nu - \nu_{J^{\prime}_{\ell} F^{\prime}_{\ell}, 
	J_{\ell} F_{\ell}} ) \; 
	{^{\beta_{\ell} L_{\ell} S I}\rho^{K_{\ell}}_0}(J^{\prime}_{\ell} 
	F^{\prime}_{\ell}, J^{\prime}_{\ell} F^{\prime}_{\ell})
	+ \varepsilon_i^{\rm coll}(\nu,\vec{\Omega}) \; ,
\label{Eq:eps_CS_2}
\end{align}
where, in order to simply the notation, we have introduced the quantity
\begin{align}
	\Upsilon(L_u L_{\ell} S I, J_u J^{\prime}_u J_{\ell} J^{\prime}_{\ell} 
	F_u F^{\prime}_u F_{\ell} F^{\prime}_{\ell}) = \, &
	(2J_u + 1)(2J^{\prime}_u + 1)(2J_{\ell} + 1)(2J^{\prime}_{\ell} + 1)
	(2F_u + 1)(2F^{\prime}_u + 1)(2F_{\ell} + 1)(2F^{\prime}_{\ell} + 1) 
	\nonumber \\
	& \times \,
	\left\{ \begin{array}{ccc}
		L_u & L_{\ell} & 1 \\
		J_{\ell} & J_u & S 
	\end{array} \right\}
	\left\{ \begin{array}{ccc}
		L_u & L_{\ell} & 1 \\
		J_{\ell} & J^{\prime}_u & S
	\end{array} \right\}
	\left\{ \begin{array}{ccc}
		L_u & L_{\ell} & 1 \\
		J^{\prime}_{\ell} & J_u & S
	\end{array} \right\}
	\left\{ \begin{array}{ccc}
		L_u & L_{\ell} & 1 \\
		J^{\prime}_{\ell} & J^{\prime}_u & S
	\end{array} \right\} \nonumber \\
	& \times \,
	\left\{ \begin{array}{ccc}
		J_u & J_{\ell} & 1 \\
		F_{\ell} & F_u & I
	\end{array} \right\}
	\left\{ \begin{array}{ccc}
		J^{\prime}_u & J_{\ell} & 1 \\
		F_{\ell} & F^{\prime}_u & I
	\end{array} \right\}
	\left\{ \begin{array}{ccc}
		J_u & J^{\prime}_{\ell} & 1 \\
		F^{\prime}_{\ell} & F_u & I
	\end{array} \right\}
	\left\{ \begin{array}{ccc}
		J^{\prime}_u & J^{\prime}_{\ell} & 1 \\
		F^{\prime}_{\ell} & F^{\prime}_u & I
	\end{array} \right\} \; .
\end{align}
As in the CRD case, the quantity $\epsilon^{\prime}$ is defined by 
\begin{equation}
	\epsilon^{\prime} = \frac{\mathcal{C}_S(L_u \rightarrow L_{\ell})}
	{A(L_u \rightarrow L_{\ell})} \; ,
\end{equation}
where $\mathcal{C}_S(L_u \rightarrow L_{\ell})$ is the inelastic collision 
rate for the transition from the upper to the lower term 
\citep[see Eq.~(21) of][and the discussion therein]{Bel15}.
The term $\varepsilon_i^{\rm coll}(\nu,\vec{\Omega})$, which represents 
the contribution to the emission coefficient brought by collisionally excited
atoms, is given by \citep[see Eq.~(37) of][]{Bel15}
\allowdisplaybreaks
\begin{align}
	\varepsilon_i^{\rm coll.}(\nu, \vec{\Omega}) = & 
	\frac{h \nu}{4 \pi} \, \mathcal{N}_T \, B(L_{\ell} \rightarrow L_u) \, 
	(2L_u + 1) (2L_{\ell} + 1) \, \epsilon^{\prime} \, B_T(\nu_0) 
	\nonumber \\
	& \times \, 
	\sum_{J_{\ell} J^{\prime}_{\ell}} \, 
	\sum_{J_u J^{\prime}_u} \, 
	\sum_{F_{\ell} F^{\prime}_{\ell}} 
	\sum_{F_u F^{\prime}_u} 
	\Upsilon(L_u L_{\ell} S I, J_u J^{\prime}_u J_{\ell} J^{\prime}_{\ell} 
	F_u F^{\prime}_u F_{\ell} F^{\prime}_{\ell}) \, \frac{1}{2} \, 
	\left[ \frac{ \Phi(\nu_{J_u F_u, J_{\ell} F_{\ell}} - \nu) + 
	\Phi(\nu_{J^{\prime}_u F^{\prime}_u, J_{\ell} F_{\ell}} - \nu)^{\ast}}
	{1 + \epsilon^{\prime} + 2 \pi {\rm i} 
	\nu_{J^{\prime}_u F^{\prime}_u, J_u F_u} / A(L_u \rightarrow L_{\ell})}
	\right] \nonumber \\
	& \times \,
	\sum_K (-1)^{F_u - F^{\prime}_u - F_{\ell} + F^{\prime}_{\ell} + K} \,
	\sqrt{3} \,
	\left\{ \begin{array}{ccc}
		K & F_u & F^{\prime}_u \\
		F_{\ell} & 1 & 1 \\
	\end{array} \right\} 
	\left\{ \begin{array}{ccc}
		F^{\prime}_u & F^{\prime}_{\ell} & 1 \\
		F^{\prime}_{\ell} & F_u & K
	\end{array} \right\} \,
	\mathcal{T}^K_0(i,\vec{\Omega}) \;
	{^{\beta_{\ell} L_{\ell} S I}\rho^K_0}(J^{\prime}_{\ell} 
	F^{\prime}_{\ell}, J^{\prime}_{\ell} F^{\prime}_{\ell}) \; ,
\label{Eq:eps_coll}
\end{align}
where $B_T(\nu_0)$ is the Planck function in the Wien limit (consistently with 
our assumption of neglecting stimulated emission), at the frequency 
$\nu_0 = (E(L_u) - E(L_{\ell}))/h$, with $E(L_u)$ and $E(L_{\ell})$ the 
energies of the centers of gravity of the upper and lower term, respectively, 
and $h$ the Planck constant.

\section{The $R_{II}$ redistribution matrix}
Recalling the definition of the radiation field tensor (see 
Eq.~(\ref{Eq:JKQ})), the emission coefficient of Eq.~(\ref{Eq:eps_CS_2}) can 
be expressed in terms of a suitable redistribution matrix \citep[$R_{II}$ 
following the terminology introduced by][]{Hum62}.
Following the convention according to which unprimed quantities refer to the 
scattered radiation while primed quantities refer to the incident radiation,
we can write
\begin{equation}
	\varepsilon_i(\nu,\vec{\Omega}) = \int {\rm d} \nu^{\prime} 
	\int \frac{{\rm d} \Omega^{\prime}}{4 \pi} \sum_{j=0}^3 
	\left[ R_{II}(\nu^{\prime}, \vec{\Omega}^{\prime}; \nu, \vec{\Omega}) 
	\right]_{ij} I_j(\nu^{\prime},\vec{\Omega}^{\prime}) + 
	\varepsilon_i^{\rm coll.}(\nu,\vec{\Omega}) \; ,
\label{Eq:emis_coef}
\end{equation}
with
\allowdisplaybreaks
\begin{align}
	\left[ R_{II}(\nu^{\prime}, \vec{\Omega}^{\prime}; \nu, \vec{\Omega}) 
	\right]_{ij} = & \frac{h \nu}{4 \pi} \, \mathcal{N}_T \, 
	B(L_{\ell} \rightarrow L_u) \, (2L_u + 1) (2L_{\ell} + 1) \nonumber \\
	& \times \, 
	\sum_{J_{\ell} J^{\prime}_{\ell}} \, 
	\sum_{J_u J^{\prime}_u} \, 
	\sum_{F_{\ell} F^{\prime}_{\ell}} 
	\sum_{F_u F^{\prime}_u} 
	\Upsilon(L_u L_{\ell} S I, J_u J^{\prime}_u J_{\ell} J^{\prime}_{\ell} 
	F_u F^{\prime}_u F_{\ell} F^{\prime}_{\ell}) \nonumber \\
	& \times \, \frac{1}{2} \, \left[ 
	\frac{ \Phi(\nu_{J_u F_u, J_{\ell} F_{\ell}} - \nu) + 
	\Phi(\nu_{J^{\prime}_u F^{\prime}_u, J_{\ell} F_{\ell}} - \nu)^{\ast}}
	{1 + \epsilon^{\prime} + 2 \pi {\rm i} 
	\nu_{J^{\prime}_u F^{\prime}_u, J_u F_u} / A(L_u \rightarrow L_{\ell})}
	\right] \, \delta(\nu - \nu^{\prime} - \nu_{J^{\prime}_{\ell} 
	F^{\prime}_{\ell}, J_{\ell} F_{\ell}}) \nonumber \\
	& \times \, 
	\sum_{K Q} \, \sum_{K_r Q_r} \, \sum_{K_{\ell}} \, 
	(-1)^{1 + K_{\ell} + F_u + F_{\ell}} \, 
	3 \sqrt{(2K+1)(2K_r +1)(2K_{\ell} +1)} 
	\nonumber \\
	& \times \,
	\left\{ \begin{array}{ccc}
		K & F_u & F^{\prime}_u \\
		F_{\ell} & 1 & 1
	\end{array} \right\} 
	\left( \begin{array}{ccc}
		K & K_r & K_{\ell} \\
		-Q & -Q_r & 0
	\end{array} \right) \,
	\left\{ \begin{array}{ccc}
		K & K_r & K_{\ell} \\
		F^{\prime}_u & 1 & F^{\prime}_{\ell} \\
		F_u & 1 & F^{\prime}_{\ell} 
	\end{array} \right\} 
	\nonumber \\
	& \times \, 
	\mathcal{T}^K_Q(i,\vec{\Omega}) \, 
	\mathcal{T}^{K_r}_{Q_r}(j,\vec{\Omega}^{\prime}) \;
	{^{\beta_{\ell} L_{\ell} S I}\rho^{K_{\ell}}_0}(J^{\prime}_{\ell} 
	F^{\prime}_{\ell}, J^{\prime}_{\ell} F^{\prime}_{\ell}) \; . 
\label{Eq:RII_atom}
\end{align}

The $R_{II}$ redistribution matrix derived above is valid in the atom rest 
frame.
In order to find the corresponding expression in the observer's frame, one 
has to take the Doppler effect into account for the given velocity distribution 
of the atoms.
The derivation is similar to that outlined in \cite{Bel14} and will not be 
given here.
Assuming that the atoms have a Maxwellian distribution of velocities, 
characterized by the temperature $T$, it can be demonstrated that the 
expression of the $R_{II}$ redistribution matrix in the observer's frame is 
obtained by performing the following substitution in Eq.~(\ref{Eq:RII_atom})
\begin{align}
	\frac{1}{2} \, & \left[ 
	\frac{ \Phi(\nu_{J_u F_u, J_{\ell} F_{\ell}} - \nu) + 
	\Phi(\nu_{J^{\prime}_u F^{\prime}_u, J_{\ell} F_{\ell}} - \nu)^{\ast}}
	{1 + \epsilon^{\prime} + 2 \pi {\rm i} 
	\nu_{J^{\prime}_u F^{\prime}_u, J_u F_u} / A(L_u \rightarrow L_{\ell})}
	\right] \, \delta(\nu - \nu^{\prime} - \nu_{J^{\prime}_{\ell} 
	F^{\prime}_{\ell}, J_{\ell} F_{\ell}}) \rightarrow \nonumber \\
	& \qquad \qquad
	\frac{1}{1 + \epsilon^{\prime} + 2 \pi {\rm i} 
	\nu_{J^{\prime}_u F^{\prime}_u, J_u F_u} / 
	A(L_u \rightarrow L_{\ell})} \, 
	\frac{1}{\pi \, \Delta \nu^2_D \, \sin{\theta}} \, 
	\exp{ \left[ - \left( \frac{\nu^{\prime} - \nu - 
	\nu_{J^{\prime}_{\ell} F^{\prime}_{\ell}, J_{\ell} F_{\ell}}}
	{2 \, \Delta \nu_D \sin{\theta/2}} \right)^2 \right] } \nonumber \\
	& \qquad \qquad \; \times \frac{1}{2} 
	\left[ W \left( \frac{a}{\cos{\theta/2}}, 
	\frac{x^{\phantom{\prime}}_{J_u F_u, J_{\ell} F_{\ell}} + 
	x^{\prime}_{J_u F_u, J^{\prime}_{\ell} F^{\prime}_{\ell}}}
	{2 \cos{\theta/2}} \right)
	+ W \left( \frac{a}{\cos{\theta/2}}, 
	\frac{x^{\phantom{\prime}}_{J^{\prime}_u F^{\prime}_u, 
	J_{\ell} F_{\ell}} + 
	x^{\prime}_{J^{\prime}_u F^{\prime}_u, J^{\prime}_{\ell} 
	F^{\prime}_{\ell}}}{2 \cos{\theta/2}} \right)^{\ast} \right] \; ,
\end{align}
where $\Delta \nu_D$ is the Doppler width in frequency units, and $\theta$ the 
scattering angle. The complex profile $W(\alpha,\beta)$ is defined as
\begin{equation}
	W(\alpha,\beta) = H(\alpha,\beta) + {\rm i} L(\alpha,\beta) \; ,
\end{equation}
with $H$ and $L$ the Voigt and Faraday-Voigt functions, respectively.
The damping parameter $a$ is given by
\begin{equation}
	a = \frac{\Gamma}{4 \pi \Delta \nu_D} \; ,
\label{Eq:adamp}
\end{equation}
with $\Gamma$ the broadening constant of the upper level (we recall that 
the lower level is assumed to be infinitely sharp). We assume that the various 
HFS levels of the upper term are characterized by the same broadening 
constant.
In the applications shown in Sect.~\ref{Sect:results}, $\Gamma$ is calculated 
including the contributions due to radiative and collisional decays from the 
upper to the lower term, and the contribution of elastic 
collisions\footnote{\label{FN:el_coll} We recall that elastic collisions play 
three different, although intimately related, roles: they contribute to the 
broadening of the spectral lines, they redistribute the photon frequency during 
the scattering process, and they relax atomic polarization. In this work, we 
heuristically take the first effect into account through the damping parameter 
$a$, but we neglect the second and third ones.}
\begin{equation}
	\Gamma = \Gamma_R + \Gamma_I + \Gamma_E = A(L_u \rightarrow L_{\ell}) 
	+ \mathcal{C}_S(L_u \rightarrow L_{\ell}) + Q_{\rm el.} \; ,
\end{equation}
with $Q_{\rm el.}$ the rate of elastic collisions.
The reduced frequencies $x_{ab}$ and $x^{\prime}_{ab}$ are given by
\begin{equation}
	x_{ab} = \frac{\nu_{ab} - \nu}{\Delta \nu_D} \; ,
	\quad {\rm and} \quad
	x^{\prime}_{ab} = \frac{\nu_{ab} - \nu^{\prime}}{\Delta \nu_D} \; .
\end{equation}
In the observer's frame, the collisional term $\varepsilon^{\rm coll.}_i(\nu,
\vec{\Omega})$ and the absorption coefficient $\eta_i(\nu,\vec{\Omega})$ are 
still given by Eqs.~(\ref{Eq:eps_coll}) and (\ref{Eq:eta_2}), respectively, 
with the only difference being that the Lorentzian and the associated 
dispersion profiles entering the definition of the complex profile 
$\Phi(\nu_0 - \nu)$ (see Eq.~(\ref{Eq:Phi_atom})) are now the Voigt and the 
Faraday-Voigt profiles, respectively.

The numerical calculation of the $R_{II}$ redistribution matrix of 
Eq.~(\ref{Eq:RII_atom}) is rather demanding since the angular and frequency 
dependencies cannot be factorized as in the atom rest frame. 
For this reason, it is customary to work with an approximate expression, 
obtained by averaging the frequency-dependent terms of the redistribution 
matrix over all the possible propagation directions $\vec{\Omega}^{\prime}$ 
and $\vec{\Omega}$ of the incoming and outgoing photons \citep[see][]{Ree82}.
Observing that this average can be easily reduced to an integral over the 
scattering angle $\theta$, the ``angle-averaged'' observer's frame expression 
of the redistribution matrix is obtained by performing the following 
substitution in Eq.~(\ref{Eq:RII_atom})
\begin{align}
	\frac{1}{2} \, & \left[ 
	\frac{ \Phi(\nu_{J_u F_u, J_{\ell} F_{\ell}} - \nu) + 
	\Phi(\nu_{J^{\prime}_u F^{\prime}_u, J_{\ell} F_{\ell}} - \nu)^{\ast}}
	{1 + \epsilon^{\prime} + 2 \pi {\rm i} 
	\nu_{J^{\prime}_u F^{\prime}_u, J_u F_u} / A(L_u \rightarrow L_{\ell})}
	\right] \, \delta(\nu - \nu^{\prime} - \nu_{J^{\prime}_{\ell} 
	F^{\prime}_{\ell}, J_{\ell} F_{\ell}}) \rightarrow \nonumber \\
	& \qquad \qquad
	\frac{1}{1 + \epsilon^{\prime} + 2 \pi {\rm i} 
	\nu_{J^{\prime}_u F^{\prime}_u, J_u F_u} / 
	A(L_u \rightarrow L_{\ell})} \, 
	\frac{1}{2 \pi \, \Delta \nu^2_D} \, \int_0^{\pi} {\rm d} \theta
	\exp{ \left[ - \left( \frac{\nu^{\prime} - \nu - 
	\nu_{J^{\prime}_{\ell} F^{\prime}_{\ell}, J_{\ell} F_{\ell}}}
	{2 \, \Delta \nu_D \sin{\theta/2}} \right)^2 \right] } \nonumber \\
	& \qquad \qquad \; \times \frac{1}{2} 
	\left[ W \left( \frac{a}{\cos{\theta/2}}, 
	\frac{x^{\phantom{\prime}}_{J_u F_u, J_{\ell} F_{\ell}} + 
	x^{\prime}_{J_u F_u, J^{\prime}_{\ell} F^{\prime}_{\ell}}}
	{2 \cos{\theta/2}} \right)
	+ W \left( \frac{a}{\cos{\theta/2}}, 
	\frac{x^{\phantom{\prime}}_{J^{\prime}_u F^{\prime}_u, 
	J_{\ell} F_{\ell}} + 
	x^{\prime}_{J^{\prime}_u F^{\prime}_u, J^{\prime}_{\ell} 
	F^{\prime}_{\ell}}}{2 \cos{\theta/2}} \right)^{\ast} \right] \; .
\end{align}

Using Eq.~(\ref{Eq:rho_lt_2}), we obtain the following final expression of the 
angle-averaged redistribution matrix
\begin{equation}
	\left[ R_{II-AA}(\nu^{\prime}, \vec{\Omega}^{\prime};\nu, \vec{\Omega}) 
	\right]_{ij} = k_L \, \sum_{KQ} \mathcal{T}^K_Q(i, \vec{\Omega}) \, 
	\sum_{K_r Q_r} \mathcal{T}^{K_r}_{Q_r}(j, \vec{\Omega}^{\prime}) \,
	r_{II-AA}(K Q, K_r Q_r; \nu^{\prime}, \nu) \; ,
\label{Eq:RII_obs}
\end{equation}
with
\allowdisplaybreaks
\begin{align}
	r_{II-AA}(K Q, K_r Q_r; \nu^{\prime}, \nu) = & \, 
	\frac{(2L_u + 1)}{(2S + 1)(2I + 1)}
	\sum_{J_{\ell} J^{\prime}_{\ell}} \, 
	\sum_{J_u J^{\prime}_u} \, 
	\sum_{F_{\ell} F^{\prime}_{\ell}} \,
	\sum_{F_u F^{\prime}_u} \,
	\sqrt{2F^{\prime}_{\ell} + 1} \,
	\Upsilon(L_u L_{\ell} S I, J_u J^{\prime}_u J_{\ell} J^{\prime}_{\ell} 
	F_u F^{\prime}_u F_{\ell} F^{\prime}_{\ell}) 
	\nonumber \\
	& \times \, 
	\frac{1}{1 + \epsilon^{\prime} + 2 \pi {\rm i} 
	\nu_{J^{\prime}_u F^{\prime}_u, J_u F_u} / 
	A(L_u \rightarrow L_{\ell})} \, 
	\frac{1}{2 \pi \, \Delta \nu^2_D} \, \int_0^{\pi} {\rm d} \theta
	\, \Bigg\{ \exp{ \left[ - \left( \frac{\nu^{\prime} - \nu - 
	\nu_{J^{\prime}_{\ell} F^{\prime}_{\ell}, J_{\ell} F_{\ell}}}
	{2 \, \Delta \nu_D \sin{\theta/2}} \right)^2 \right] }  
	\nonumber \\
	& \times \, 
	\frac{1}{2} \left[ W \left( \frac{a}{\cos{\theta/2}}, 
	\frac{x^{\phantom{\prime}}_{J_u F_u, J_{\ell} F_{\ell}} + 
	x^{\prime}_{J_u F_u, J^{\prime}_{\ell} F^{\prime}_{\ell}}}
	{2 \cos{\theta/2}} \right)
	+ W \left( \frac{a}{\cos{\theta/2}}, 
	\frac{x^{\phantom{\prime}}_{J^{\prime}_u F^{\prime}_u, 
	J_{\ell} F_{\ell}} + 
	x^{\prime}_{J^{\prime}_u F^{\prime}_u, J^{\prime}_{\ell} 
	F^{\prime}_{\ell}}}{2 \cos{\theta/2}} \right)^{\ast} \right] \Bigg\}
	\nonumber \\
	& \times \, 
	\sum_{K_{\ell}} \,
	(-1)^{1 + K_{\ell} + F_u + F_{\ell}} \, 
	3 \sqrt{(2K +1)(2K_r +1)(2K_{\ell} +1)}
	\nonumber \\
	& \times \,
	\left\{ \begin{array}{ccc}
		K & F_u & F^{\prime}_u \\
		F_{\ell} & 1 & 1
	\end{array} \right\} 
	\left( \begin{array}{ccc}
		K & K_r & K_{\ell} \\
		-Q & -Q_r & 0
	\end{array} \right) \,
	\left\{ \begin{array}{ccc}
		K & K_r & K_{\ell} \\
		F^{\prime}_u & 1 & F^{\prime}_{\ell} \\
		F_u & 1 & F^{\prime}_{\ell} 
	\end{array} \right\} \,
	\sigma^{K_{\ell}}_0(J^{\prime}_{\ell}, F^{\prime}_{\ell}) \; .
\end{align}
As a consistency check, we verified that neglecting lower term polarization 
(i.e., setting $\sigma^{K_{\ell}}_0(J^{\prime}_{\ell}, F^{\prime}_{\ell}) = 
\delta_{K_{\ell} 0})$, and in the absence of HFS (i.e., setting $I=0$), 
Eq.~(\ref{Eq:RII_obs}) reduces to Eq.~(35) of \citet{Bel14}.
Likewise, substituting Eq.~(\ref{Eq:rho_lt_2}) into Eq.~(\ref{Eq:eps_coll}), 
the collisional term can be written in the form
\begin{equation}
	\varepsilon_i^{\rm coll.}(\nu, \vec{\Omega}) = k_L \, 
	\sum_K \mathcal{T}^K_0(i,\vec{\Omega}) \, B_T(\nu_0) \, 
	\beta^K_0(\nu) \; ,
\label{Eq:eps_coll_obs}
\end{equation}
with
\allowdisplaybreaks
\begin{align}
	\beta^K_0(\nu) = & \, 
	\frac{(2L_u + 1)}{(2S + 1)(2I + 1)}
	\sum_{J_{\ell} J^{\prime}_{\ell}} \,
	\sum_{J_u J^{\prime}_u} \,
	\sum_{F_{\ell} F^{\prime}_{\ell}} \, 
	\sum_{F_u F^{\prime}_u} \,
	\sqrt{2F^{\prime}_{\ell} +1} \,
	\Upsilon(L_u L_{\ell} S I, J_u J^{\prime}_u J_{\ell} J^{\prime}_{\ell} 
	F_u F^{\prime}_u F_{\ell} F^{\prime}_{\ell}) 
	\nonumber \\
	& \times \, \frac{1}{2} \, 
	\left[ \frac{ \Phi(\nu_{J_u F_u, J_{\ell} F_{\ell}} - \nu) + 
	\Phi(\nu_{J^{\prime}_u F^{\prime}_u, J_{\ell} F_{\ell}} - \nu)^{\ast}}
	{1 + \epsilon^{\prime} + 2 \pi {\rm i} 
	\nu_{J^{\prime}_u F^{\prime}_u, J_u F_u} / A(L_u \rightarrow L_{\ell})}
	\right] \, \epsilon^{\prime}
	\nonumber \\
	& \times \, 
	\sum_K (-1)^{F_u - F^{\prime}_u - F_{\ell} + F^{\prime}_{\ell} + K} \,
	\sqrt{3} 
	\left\{ \begin{array}{ccc}
		K & F_u & F^{\prime}_u \\
		F_{\ell} & 1 & 1 
	\end{array} \right\} 
	\left\{ \begin{array}{ccc}
		F^{\prime}_u & F^{\prime}_{\ell} & 1 \\
		F^{\prime}_{\ell} & F_u & K
	\end{array} \right\} \,
	\sigma^K_0(J^{\prime}_{\ell}, F^{\prime}_{\ell}) \; .
\end{align}
Also in this case, we verified that if lower term polarization is neglected 
and in the absence of HFS, Eq.~(\ref{Eq:eps_coll_obs}) reduces to Eq.~(29) of 
\citet{Bel14}.

\section{The radiative transfer equations}
We consider a plane-parallel atmosphere in the absence of magnetic fields.
The problem is thus characterized by cylindrical symmetry along the direction
perpendicular to the surface of the atmosphere (hereafter, the ``vertical'').
Taking the quantization axis directed along the vertical, and the reference 
direction for positive $Q$ parallel to the surface, at any height in the 
atmosphere the radiation field is described by the Stokes parameters $I$ and 
$Q$ only (hereafter always indicated as $I_0$ and $I_1$, respectively), while 
$J^0_0$ and $J^2_0$ are the only non-vanishing elements of the radiation field 
tensor.
Under the above-mentioned hypotheses, and neglecting stimulated emission, the 
RT equation takes the form
\begin{equation}
	\frac{\rm d}{{\rm d} s} 
	\left( \begin{array}{c}
		I_0(\nu, \vec{\Omega}) \\
		I_1(\nu, \vec{\Omega}) 
	\end{array} \right)
	= - \left( \begin{array}{cc}
		\eta_0(\nu, \vec{\Omega}) & \eta_1(\nu,\vec{\Omega}) \\
		\eta_1(\nu, \vec{\Omega}) & \eta_0(\nu,\vec{\Omega})
	\end{array} \right)
	\left( \begin{array}{c}
		I_0(\nu, \vec{\Omega}) \\
		I_1(\nu, \vec{\Omega}) 
	\end{array} \right) +
	\left( \begin{array}{c}
		\varepsilon_0(\nu, \vec{\Omega}) \\
		\varepsilon_1(\nu, \vec{\Omega}) 
	\end{array} \right) \; ,
\label{Eq:RT1}
\end{equation}
where $s$ is the spatial coordinate measured along the ray path.
The emission and absorption coefficients appearing in Eq.~(\ref{Eq:RT1}) 
contain both the line and the continuum contributions (hereafter indicated 
with the apices $\ell$ and $c$, respectively):
\begin{align}
	\eta_i(\nu,\vec{\Omega}) & = \eta^{\ell}_i(\nu,\vec{\Omega}) + 
	\eta^{c}_i(\nu,\vec{\Omega}) \; , \\
	\varepsilon_i(\nu,\vec{\Omega}) & = 
	\varepsilon^{\ell}_i(\nu,\vec{\Omega}) + 
	\varepsilon^{c}_i(\nu,\vec{\Omega}) \; .
\end{align}
The line absorption coefficient, $\eta^{\ell}_i(\nu, \vec{\Omega})$, is given 
by Eq.~(\ref{Eq:eta_2}). 
Using Eqs.~(\ref{Eq:emis_coef}), (\ref{Eq:RII_obs}), and 
(\ref{Eq:eps_coll_obs}), the line emission coefficient, 
$\varepsilon^{\ell}_i(\nu, \vec{\Omega})$, can be written in the form
\begin{equation}
	\varepsilon^{\ell}_i(\nu,\vec{\Omega}) = k_L \sum_{KQ} 
	\mathcal{T}^K_Q(i,\vec{\Omega}) \, \left( \tilde{J}^K_Q(\nu) + 
	\delta_{Q0} \, B_T(\nu_0) \, \beta^K_0(\nu) \right) \; ,
\label{Eq:epsl_F}
\end{equation}
with
\begin{equation}
	\tilde{J}^K_Q(\nu) = \sum_{K_r Q_r} \int {\rm d} \nu^{\prime} 
	J^{K_r}_{Q_r}(\nu^{\prime}) \, r_{II-AA}(K Q, K_r Q_r; \nu^{\prime}, 
	\nu) \; .
\end{equation}

We consider the contribution of a coherent polarized continuum.
Neglecting dichroism (which is a very good approximation in the visible part of 
the solar spectrum), the continuum total absorption coefficient (opacity) is 
given by
\begin{equation}
	\eta^c_i(\nu) = [ k_c(\nu) + \sigma(\nu) ] \, \delta_{i0} \; ,
\end{equation}
with $k_c(\nu)$ the continuum true absorption coefficient, and $\sigma(\nu)$ 
the continuum scattering coefficient.
The continuum emission coefficient is given by
\begin{equation}
	\varepsilon^c_i(\nu,\vec{\Omega}) = \sigma(\nu) \sum_{KQ} 
	\mathcal{T}^K_Q(i,\vec{\Omega}) \, (-1)^Q \, J^K_{-Q}(\nu) \,
	+ \varepsilon^c_{\rm th}(\nu) \, \delta_{i 0} \; .
\label{Eq:epsc_F}
\end{equation}
The first term in the right-hand side of Eq.~(\ref{Eq:epsc_F}) represents 
the contribution to the continuum emission coefficient coming from coherent 
scattering processes (Rayleigh and Thomson scattering), the second term 
represents the thermal contribution (which does not contribute to the 
polarization of the continuum). 
Under the assumption that the continuum is in LTE, $\varepsilon^c_{\rm th}(\nu) 
= k_c(\nu) \, B_T(\nu)$.

As is clear from Eq.~(\ref{Eq:RT1}), when dichroism is taken into account 
(i.e., when $\eta_1$ is non-zero), the RT equations for the Stokes parameters 
$I_0$ and $I_1$ are coupled to each other.
However, they can be easily decoupled by introducing the quantities
\begin{align}
	\hat{I}_0(\nu,\vec{\Omega}) = & \, I_0(\nu,\vec{\Omega}) + 
	I_1(\nu,\vec{\Omega}) \; , \nonumber \\
	\hat{I}_1(\nu,\vec{\Omega}) = & \, I_0(\nu,\vec{\Omega}) -
	I_1(\nu,\vec{\Omega}) \; .
\label{Eq:Icap}
\end{align}
Indeed, from Eqs.~(\ref{Eq:RT1}) and (\ref{Eq:Icap}), it can be easily shown 
that the RT equations for $\hat{I}_j(\nu,\vec{\Omega})$ ($j=0,1$) are given by
\begin{equation}
	\frac{\rm d}{{\rm d} s} \, \hat{I}_j(\nu, \vec{\Omega}) = 
	- \hat{\eta}_j(\nu, \vec{\Omega}) \, \hat{I}_j(\nu,\vec{\Omega}) +
	\hat{\varepsilon}_j(\nu, \vec{\Omega}) \; ,
\label{Eq:RT2}
\end{equation}
with
\begin{align}
	\hat{\varepsilon}_j(\nu,\vec{\Omega}) & = 
	\varepsilon_0(\nu,\vec{\Omega}) + (-1)^j \,
	\varepsilon_1(\nu,\vec{\Omega}) \; , \\ 
	\hat{\eta}_j(\nu,\vec{\Omega}) & =
	\eta_0(\nu,\vec{\Omega}) + (-1)^j \,
	\eta_1(\nu,\vec{\Omega}) \; .
\end{align}
The Stokes parameters $I_0$ and $I_1$ can be easily obtained from the new 
quantities through the relations
\begin{align}
	I_0(\nu,\vec{\Omega}) = & \, \frac{1}{2} \left( 
	\hat{I}_0(\nu,\vec{\Omega}) + \hat{I}_1(\nu,\vec{\Omega}) \right) \; , 
	\nonumber \\
	I_1(\nu,\vec{\Omega}) = & \, \frac{1}{2} \left( 
	\hat{I}_0(\nu,\vec{\Omega}) - \hat{I}_1(\nu,\vec{\Omega}) \right) \; .
\label{Eq:Istd}
\end{align}
It can be easily shown that the quantities 
$\hat{\eta}^{\ell}_j(\nu,\vec{\Omega})$, and 
$\hat{\varepsilon}^{\ell}_j(\nu,\vec{\Omega})$ 
are given by Eqs.~(\ref{Eq:eta_2}) and (\ref{Eq:epsl_F}), respectively, 
provided that the geometrical tensor $\mathcal{T}^K_Q(i,\vec{\Omega})$ 
appearing in such equations is substituted by
\begin{equation}
	\hat{\mathcal{T}}^K_Q(j,\vec{\Omega}) = 
	\mathcal{T}^K_Q(0,\vec{\Omega}) + 
	(-1)^j \, \mathcal{T}^K_Q(1,\vec{\Omega}) \qquad j=0,1 \; .
\end{equation}
On the other hand, we have
\begin{equation}
	\hat{\eta}^c_j(\nu) = \eta^c_0(\nu)= k_c(\nu)+ \sigma(\nu)
	\qquad j=0,1 \; ,
\end{equation}
and
\begin{equation}
	\hat{\varepsilon}^c_j(\nu,\vec{\Omega}) = \sigma(\nu) \sum_{KQ} 
	\hat{\mathcal{T}}^K_Q(j,\vec{\Omega}) \, (-1)^Q \, J^K_{-Q}(\nu) \,
	+ \varepsilon^c_{\rm th}(\nu) \qquad j=0,1 \; .
\end{equation}
In terms of the new quantities $\hat{I}_j(\nu,\vec{\Omega})$ and 
$\hat{\mathcal{T}}^K_Q(j,\vec{\Omega})$, the radiation field tensor is given by
\begin{equation}
	J^K_Q(\nu) = \int \frac{{\rm d} \Omega}{4 \pi} \sum_{j=0}^1 
	\mathcal{T}^K_Q(j,\vec{\Omega}) \, I_j(\nu,\vec{\Omega}) 
	= \frac{1}{2} \int \frac{{\rm d} \Omega}{4 \pi} \sum_{j=0}^1 
	\hat{\mathcal{T}}^K_Q(j,\vec{\Omega}) \, \hat{I}_j(\nu,\vec{\Omega}) 
	\; .
\label{Eq:JKQ2}
\end{equation}

Introducing the optical depth $\hat{\tau}_j(\nu,\vec{\Omega})$ defined by
\begin{equation}
	{\rm d} \hat{\tau}_j(\nu,\vec{\Omega}) = 
	-\hat{\eta}_j(\nu,\vec{\Omega}) \, {\rm d}s \; ,
\end{equation}
and the source function 
\begin{equation}
	\hat{S}_{\!\! j}(\nu,\vec{\Omega}) = 
	\frac{\hat{\varepsilon}_j(\nu,\vec{\Omega})}
	{\hat{\eta}_j(\nu,\vec{\Omega})} \; ,
\end{equation}
the RT equation for the quantities $\hat{I}_j$ takes the form
\begin{equation}
	\frac{\rm d}{{\rm d} \hat{\tau}_j(\nu,\vec{\Omega})}
	\hat{I}_j(\nu,\vec{\Omega}) = \hat{I}_j(\nu,\vec{\Omega}) 
	- \hat{S}_{\!\! j}(\nu,\vec{\Omega}) \; .
\label{Eq:RT3}
\end{equation}
The source function $\hat{S}_{\!\! j}(\nu,\vec{\Omega})$ can be conveniently 
written in the form
\begin{equation}
	\hat{S}_{\!\! j}(\nu,\vec{\Omega}) = 
	\frac{k_L}{\hat{\eta}_j(\nu,\vec{\Omega})} \, 
	\sum_{KQ} \hat{\mathcal{T}}^K_Q(j,\vec{\Omega}) \, S^K_Q(\nu) \; ,
\end{equation}
with
\begin{equation}
	S^K_Q(\nu) = \check{J}^K_Q(\nu) + \delta_{Q 0} \, B_T(\nu_0) \, 
	\beta^K_0(\nu) + \delta_{K 0} \, \delta_{Q 0} \, 
	\frac{\varepsilon^c_{\rm th}(\nu)}{k_L} \; .
\label{Eq:SKQ}
\end{equation}
The quantity $\check{J}^K_Q(\nu)$ is given by
\begin{equation}
	\check{J}^K_Q(\nu) = \tilde{J}^K_Q(\nu) + s(\nu) \, (-1)^Q 
	J^K_{-Q}(\nu) = \sum_{K_r Q_r} \int {\rm d} \nu^{\prime} 
	J^{K_r}_{Q_r}(\nu^{\prime}) \, 
	\psi_{K Q, K_r Q_r}(\nu^{\prime}, \nu) \; ,
\label{Eq:JKQ_hat}
\end{equation}
with $s(\nu) = \sigma(\nu)/k_L$, and  
\begin{equation}
	\psi_{K Q, K_r Q_r}(\nu^{\prime}, \nu) = 
	\left[ r_{II-AA}(K Q,K_r Q_r; \nu^{\prime}, \nu) + 
	\delta_{K_r K} \, \delta_{Q_r, -Q} \, \delta(\nu - \nu^{\prime}) 
	s(\nu) \, (-1)^Q \, \right] \; .
\end{equation}

\section{Iterative solution of the non-LTE problem}
The calculation of $J^0_0(\nu)$ and $J^2_0(\nu)$ at each height in the 
atmosphere requires the knowledge of the quantities 
$\hat{I}_j(\nu,\vec{\Omega})$ along the various directions of the chosen 
angular quadrature, as obtained from the solution of Eq.~(\ref{Eq:RT3}).
The formal solution of this equation is given by
\begin{equation}
	\hat{I}_j(\nu,\mu;{\rm O}) = \hat{I}_j(\nu,\mu;{\rm M}) \, 
	{\rm e}^{-\Delta \hat{\tau}_j(\nu,\mu)} + 
	\int_0^{\Delta \hat{\tau}_j(\nu,\mu)} 
	\!\! \hat{S}_{\! j}(\nu,\mu;t) \, {\rm e}^{-t} \, {\rm d}t \; ,
\label{Eq:RT_formal}
\end{equation}
where $O$ is a given height point in the considered discretization of the 
atmosphere, and $M$ is the corresponding ``upwind'' point.
As far as the dependence on the propagation direction is concerned, we have 
taken into account that, due to the cylindrical symmetry of the problem, the 
various quantities only depend on $\mu = \cos{\theta}$, with $\theta$ the angle 
between the vertical and the propagation direction.
The quantity $\Delta \hat{\tau}_j(\nu,\mu)$ is the optical distance between 
$O$ and $M$, at frequency $\nu$, along the direction specified by $\mu$.
We evaluate the integral in the right-hand side of Eq.~(\ref{Eq:RT_formal})
by means of the short-characteristic method \citep[see][]{Kun88}.
Indicating the values of $\hat{I}_j$ and $\hat{S}_j$ at the various points of 
the spatial grid through the elements of column vectors, the formal solution of 
the RT equation can be written in the form
\begin{equation}
	\hat{I}_j(\nu,\mu;\ell) = \sum_{m=1}^{N} \Lambda^j_{\nu \mu}(\ell,m) \, 
	\hat{S}_{\!\! j}(\nu,\mu;m) + \hat{T}_{\!\! j}(\nu,\mu;\ell) \; ,
\label{Eq:RT_lambda}
\end{equation}
where $\ell, m= 1, \dots, N$, with $N$ number of grid points, 
$\hat{T}_{\!\! j}(\nu,\mu; \ell)$ is the value of the transmitted 
$\hat{I}_j(\nu,\mu)$ at point $\ell$ due to the radiation incident at the 
boundary, and $\Lambda^j_{\nu \mu}(\ell,m)$ are the elements of a 
$N \times N$ operator.

Substituting Eq.~(\ref{Eq:RT_lambda}) into Eq.~(\ref{Eq:JKQ2}) the operations 
required for the numerical calculation of $J^K_0(\nu)$ at point $\ell$ can be 
indicated as follows:
\begin{equation}
	J^K_0(\nu;\ell) = \sum_{m=1}^N \sum_{K^{\prime}} 
	\Lambda_{K 0,K^{\prime} 0}(\nu; \ell, m) \, 
	S^{K^{\prime}}_0(\nu;m) + T^K_0(\nu;\ell) \; ,
\label{Eq:JKQ_SKQ}
\end{equation}
where we have explicitly indicated the dependence on the height point in the 
atmosphere of the various physical quantities previously introduced. 
The $\Lambda$ operators and the $T^K_0$ tensor are given by
\allowdisplaybreaks
\begin{align}
	& \Lambda_{K 0,K^{\prime} 0}(\nu;\ell,m) = \frac{1}{2}
	\int \frac{{\rm d} \vec{\Omega}}{4 \pi} 
	\sum_{j=0}^1 \, \Lambda^j_{\nu \mu}(\ell,m) \, 
	\frac{k_L(m)}{\hat{\eta}_j(\nu,\mu;m)} \,
	\hat{\mathcal{T}}^K_0(j,\vec{\Omega}) \,
	\hat{\mathcal{T}}^{K^{\prime}}_0(j,\vec{\Omega}) 
	\label{Eq:Lam_KQ} \; , \\
	& T^K_0(\nu;\ell) = \frac{1}{2} \int \frac{{\rm d} \vec{\Omega}}{4 \pi} 
	\, \sum_{j=0}^1 \hat{\mathcal{T}}^K_0(j,\vec{\Omega}) \, 
	\hat{T}_{\! j}(\nu,\mu;\ell) 
	\label{Eq:T_KQ} \; .
\end{align}

The equations for $S^0_0$ and $S^2_0$ resulting from the substitution of 
Eq.~(\ref{Eq:JKQ_SKQ}) into Eq.~(\ref{Eq:SKQ}) (through Eq.~(\ref{Eq:JKQ_hat})) 
represent the fundamental equations for the non-LTE problem under 
consideration. 
It is well known that the most suitable approach for the numerical solution of 
this set of equations is through iterative methods.
In this work, we apply the Jacobian-based iterative method.

Let $S_0^{0\,\rm old}$ and $S_0^{2\,\rm old}$ be given estimates of the 
unknowns at the various points of the grid. 
At any grid point $\ell$, we calculate $J^0_0$ and $J^2_0$ through 
Eq.~(\ref{Eq:JKQ_SKQ}) by using such ``old'' values of the source function at 
all the grid points, except at point $\ell$ where the new estimates, 
$S_0^{0\,\rm new}$ and $S_0^{2\,\rm new}$, are {\it implicitly} (their value 
being still unknown) used:
\begin{equation}
	J^K_0(\nu;\ell) = J^K_0(\nu; \ell)^{\rm old} + 
	\sum_{K^{\prime}} 
	\Lambda_{K 0, K^{\prime} 0}(\nu; \ell, \ell) \, 
	\Delta S^{K^{\prime}}_0(\nu; \ell) \; ,
\end{equation}
with 
\begin{equation}
	\Delta S^{K^{\prime}}_0(\nu; \ell) = 
	S^{K^{\prime}}_0(\nu;\ell)^{\rm new} -
	S^{K^{\prime}}_0(\nu;\ell)^{\rm old} \; ,
\end{equation}
and where $J^0_0(\nu; \ell)^{\rm old}$ and $J^2_0(\nu; \ell)^{\rm old}$ are 
the values of $J^0_0$ and $J^2_0$ that are obtained from a formal solution of 
the RT equation, carried out using the ``old'' estimates $S_0^{0\,\rm old}$ 
and $S_0^{2\,\rm old}$.
The following step is to calculate $\check{J}^K_0(\nu; \ell)$ through 
Eq.~(\ref{Eq:JKQ_hat}):
\begin{equation}
	\check{J}^K_0(\nu; \ell) = \check{J}^K_0(\nu; \ell)^{\rm old} + 
	\sum_{K_r K^{\prime}_r} \int {\rm d} \nu^{\prime}
	\psi_{K 0, K_r 0}(\nu^{\prime}, \nu; \ell) \,
	\Lambda_{K_r 0, K^{\prime}_r 0}(\nu^{\prime}; \ell, \ell) \, 
	\Delta S^{K^{\prime}_r}_0(\nu^{\prime}; \ell) \; .
\label{Eq:JKQ_hat_jac}
\end{equation}
The new values of the source function are finally obtained by substituting 
Eq.~(\ref{Eq:JKQ_hat_jac}) into Eq.~(\ref{Eq:SKQ}):
\begin{equation}
	\Delta S^K_0(\nu;\ell) = \sum_{K_r K^{\prime}_r} 
	\int {\rm d} \nu^{\prime} \psi_{K 0, K_r 0}(\nu^{\prime}, \nu; \ell) \, 
	\Lambda_{K_r 0, K^{\prime}_r 0}(\nu^{\prime}; \ell, \ell) \, 
	\Delta S^{K^{\prime}_r}_0(\nu^{\prime}; \ell) + R^K_0(\nu, \ell) \; ,
\end{equation}
with
\begin{equation}
	R^K_0(\nu; \ell) = \check{J}^K_0(\nu;\ell)^{\rm old} + 
	B_T(\nu_0;\ell) \, \beta^K_0(\nu; \ell) +
	\delta_{K 0} \, \frac{\varepsilon^c_{\rm th}(\nu;\ell)}{k_L(\ell)}
	- S^K_0(\nu; \ell)^{\rm old} \; .
\end{equation}

The Jacobi-based method that we apply in this work is obtained by setting 
$\Lambda_{K0,K^{\prime} 0} = \Lambda_{00,00} \, \delta_{K0} \,
\delta_{K^{\prime} 0}$ \citep[see][for a detailed discussion on the role of 
the various $\Lambda$-operators]{JTB99}.
Introducing the ensuing expressions into Eq.~(\ref{Eq:SKQ}) we obtain
\begin{align}
	\Delta S^0_0(\nu;\ell) = & \int {\rm d} \nu^{\prime}
	\psi_{00,00}(\nu^{\prime},\nu;\ell) \, 
	\Lambda_{00,00}(\nu^{\prime};\ell,\ell) \,
	\Delta S^0_0(\nu^{\prime};\ell) + R^0_0(\nu; \ell) \; ,
	\label{Eq:DeltaS00} \\
	\Delta S^2_0(\nu;\ell) = & \int {\rm d} \nu^{\prime}
	\psi_{20,00}(\nu^{\prime},\nu;\ell) \, 
	\Lambda_{00,00}(\nu^{\prime};\ell,\ell) \,
	\Delta S^0_0(\nu^{\prime};\ell) + R^2_0(\nu; \ell) \; .
	\label{Eq:DeltaS20}
\end{align}
We calculate the new estimate of $S^0_0$ by solving Eq.~(\ref{Eq:DeltaS00}) 
through the so-called ``Frequency-by-frequency'' method.

\section{Results}
\label{Sect:results}
In this section, we present and discuss the scattering polarization profiles 
of the Na~{\sc i} D$_1$ line resulting from the numerical solution of the 
non-LTE problem described in the previous sections, in two semi-empirical 
models of the solar atmosphere.
In particular, we show how the $Q/I$ signal produced in the core of the D$_1$ 
line by the mechanism pointed out by \citet{Bel13b} is affected by quantum 
interference between the upper $J$-levels of D$_1$ and D$_2$, and by the 
presence of an increasing amount of atomic polarization in the ground level of 
sodium.
Although our work is focused on the D$_1$ line, we present and discuss also 
the polarization profiles of the D$_2$ line, which is naturally included in 
our calculations.

\subsection{Sensitivity to the atmospheric model (No Lower Level Polarization)}
\begin{figure}
\centering
\includegraphics[width=0.9\textwidth]{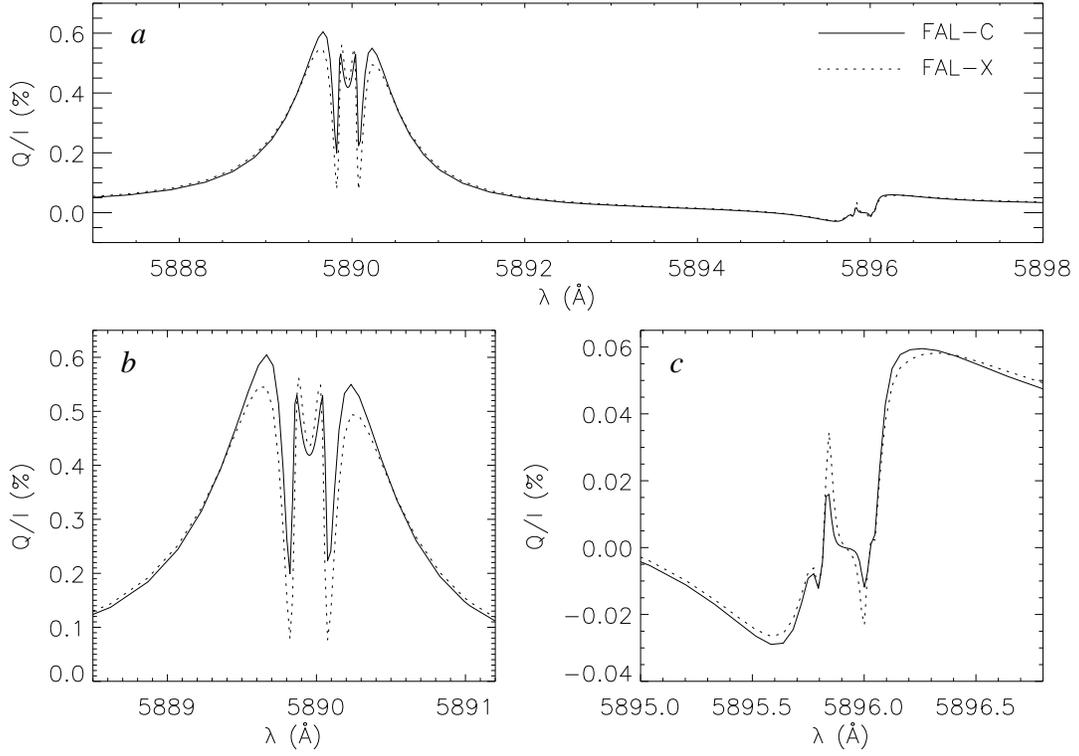}
\caption{Fractional linear polarization profiles calculated in the FAL-C 
(solid) and FAL-X (dotted) atmospheric models, for a line of sight with 
$\mu=0.1$ ($\mu$ being the cosine of the heliocentric angle), in the absence 
of lower level polarization. The reference direction for positive $Q$ is 
parallel to the solar limb.
{\it Panel a:} spectral interval containing both D$_1$ and D$_2$. 
{\it Panel b:} zoom on the D$_2$ line. 
{\it Panel c:} zoom on the D$_1$ line.}
\label{Fig:FALCX}
\end{figure}

Figure~\ref{Fig:FALCX} shows the $Q/I$ profiles calculated in model C of 
\citet{Fon93} (hereafter, FAL-C), and in model M$_{\rm CO}$ (also known as 
FAL-X) of \citet{Avr95}, assuming that there is no atomic polarization in the 
lower levels.
As far as the D$_2$ line is concerned, the theoretical profile shows the 
typical triplet peak structure that is observed in this line. 
The central peak of the calculated profile has a small but appreciable 
sub-structure, which is sensibly reduced once the $I$ and $Q$ profiles are 
convolved with a gaussian, so to take the spectral smearing due to the finite 
bandwidth of the instrument into account.
While in the observed profiles the central peak is higher than the wing peaks, 
in our theoretical profile the three peaks have approximately the same 
amplitude.
Although the presence of lower level polarization produces an increase of the 
amplitude of the central peak (see Sect.~\ref{Sect:llp}), we believe that the 
main reason for this disagreement with the observations has to be sought in the 
assumption of purely coherent scattering in the atom rest frame.
Indeed, calculations carried out within the framework of the PRD approach of 
\citet{Bom97}, considering a simpler two-level model atom, show that the 
$R_{III}$ part of the redistribution matrix, which describes the contribution 
of scattering processes in the limit of CRD, produces a decrease of the 
amplitude of the wing peaks, leaving almost unaffected the central one. 
This is not surprising since the central peak forms much higher in the 
atmosphere, where the impact of collisions capable of redistributing the photon 
frequency during the scattering process is negligible.
We finally observe that the two peaks in the wings of the line do not have the 
same amplitude, the red one being slightly smaller than the blue one. 
This is due to the effect of $J$-state interference, and it is in agreement 
with the observed profiles. 

Moving toward longer wavelengths, the theoretical profile reproduces the sign 
reversal that is observed between D$_1$ and D$_2$ very well, as well as the 
general pattern that is observed in the wings of the D$_1$ line.
We recall that these are the typical signatures of $J$-state interference
\citep[see][]{Ste80,Lan04,Bel11}, which are fully accounted for in our 
theoretical approach.
In the core of the D$_1$ line, we obtain a clear $Q/I$ signal, with 
positive and negative peaks leading to an almost null integrated 
polarization signal. 
The positive peak is slightly blue-shifted with respect to line center (where 
the signal is zero), while the negative one is slightly red-shifted. 
A small negative dip can also be recognized between the positive peak and the 
negative minimum of the $J$-state interference pattern.
Remarkably, this signal is not due either to lower level polarization or to the 
presence of a magnetic field (both ingredients have been neglected in the 
calculations of Fig.~\ref{Fig:FALCX}), but it is produced by the physical 
mechanism identified and discussed by \citet{Bel13b}.
It is important to note that this signal appears in the core of the D$_1$ line, 
where the assumption of purely coherent scattering in the atom rest frame is 
a good approximation and the contribution of $R_{III}$ (not included in this 
work) can be safely neglected.
The amplitude of the positive and negative peaks is quite sensitive to the 
atmospheric model. In particular, in agreement with the results of 
\citet{Bel13b}, the peaks obtained in FAL-X are almost two times larger than 
those calculated in FAL-C.
Nonetheless, the amplitude of our theoretical profiles remains sensibly smaller 
than that of the signal observed by \citet{Ste97}, although it is in 
agreement with other observations \citep[e.g.,][]{JTB01}.

The presence of non-zero signals in the core of the Na~{\sc i} D$_1$ line has 
been confirmed by recent observations carried out with the Z\"urich Imaging 
Polarimeter (ZIMPOL) at the Istituto Ricerche Solari Locarno (such observations 
will be published in a forthcoming paper). These signals, which show 
conspicuous variations along the slit, are much more similar to our 
theoretical profiles than to the large signal observed by \citet{Ste97}.
As it will be shown below, the presence of ad-hoc amounts of atomic 
polarization in the ground level of sodium allows us to significantly increase 
the amplitude of our theoretical profiles. 
On the other hand, the physical mechanism pointed out by \citet{Bel13b}, 
possibly together with the presence of some lower level polarization, appears 
to be perfectly suitable for explaining the $Q/I$ signals revealed by the 
recent observations.

\subsection{The impact of lower level polarization}
\label{Sect:llp}
We include now a given amount of atomic polarization in the lower $F$-levels, 
parametrizing the quantity $\sigma^2_0$ defined in Eq.~(\ref{Eq:sigmaK0}) 
according to the following expression
\begin{equation}
	\sigma^2_0(F_{\ell},h) = \frac{a(F_{\ell})}{1 + b
	(F_{\ell}) \, \tau_{\nu_0}(h)} \; ,
\end{equation}
where $h$ is the geometrical height in the atmosphere, and $\tau_{\nu_0}$ is 
the optical depth along the vertical, at the line center frequency of the D$_2$ 
line.
The parameter $a$ represents the value of $\sigma^2_0$ at the top of the 
atmosphere, while $b$ sets its scale height. 
A similar parametrization of lower level polarization was used by 
\citet{Lan98}.

\begin{figure}
\centering
\includegraphics[width=0.9\textwidth]{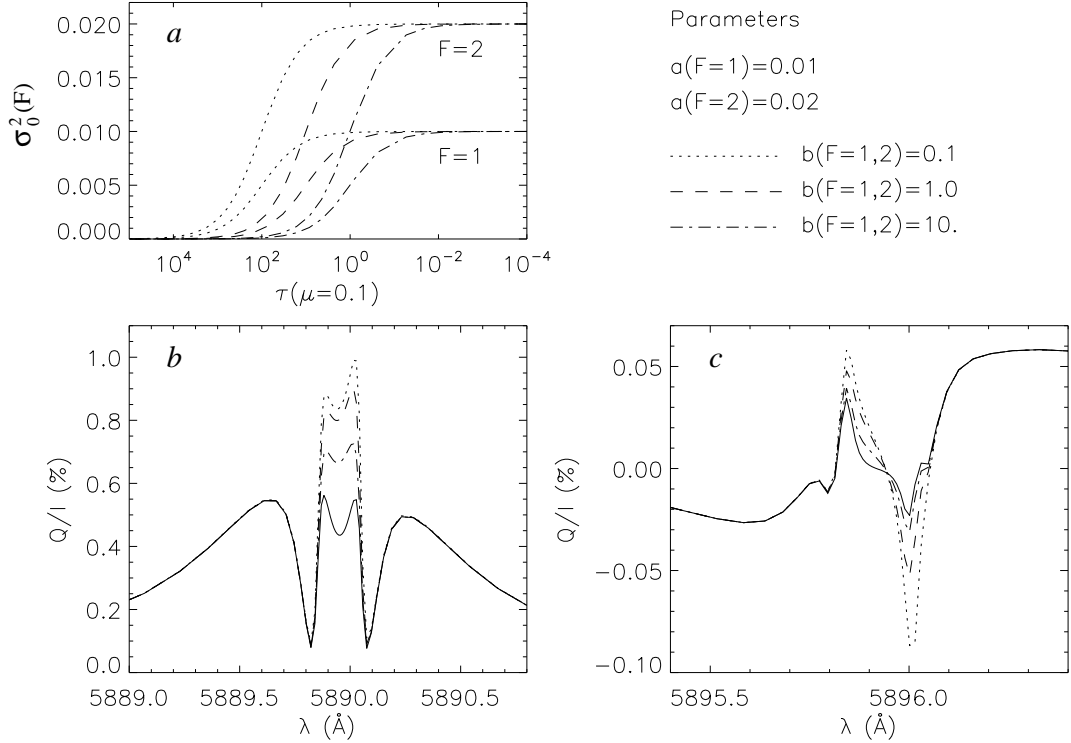}
\caption{{\it Panel a:} plot of $\sigma^2_0(F_{\ell}=2)$ (upper lines) and 
$\sigma^2_0(F_{\ell}=1)$ (lower lines) as a function of the optical depth, 
$\tau$, measured along the line of sight ($\mu=0.1$), at the line center 
frequency of the D$_2$ line, in the FAL-X model, for three different values of 
the parameter $b$.
The values of the parameters $a$ and $b$ in the three cases considered are 
indicated in the figure.
{\it Panel b:} fractional linear polarization profiles of the D$_2$ line, 
calculated for the three different parametrizations of lower level 
polarization shown in panel $a$.
The profile with the solid line corresponds to the case of no lower level 
polarization (it coincides with the dotted profile in panel $b$ of Fig.~1), 
and it is included for reference.
All of the profiles have been calculated in the FAL-X model, for $\mu=0.1$.
{\it Panel c:} same as panel $b$, but for the D$_1$ line.}
\label{Fig:llp_1}
\end{figure}

Figure~\ref{Fig:llp_1} shows the $Q/I$ profiles of the D$_1$ and D$_2$ lines 
calculated in the FAL-X model, for different values of the parameter 
$b(F_{\ell})$, keeping fixed the parameter $a(F_{\ell})$.
We set $b(F_{\ell}=1)=b(F_{\ell}=2)$, and $a(F_{\ell}=2)=2a(F_{\ell}=1)$ 
\citep[we recall that a ratio of about 2 between the atomic polarization of the 
levels $F_{\ell}=2$ and $F_{\ell}=1$ was also considered by][in order to obtain 
his best fit to the observed profiles]{Lan98}.
In the D$_2$ line, lower level polarization produces an appreciable increase 
of the amplitude of the central peak, and a slight modification of its 
sub-structure.
On the other hand, the two peaks in the wings, as well as the dips between 
the central and wing peaks are unchanged.
As expected, the increase of the central peak is larger for smaller values of 
the parameter $b$, that is, when the value of $\sigma^2_0$ starts decreasing 
below the height of formation of the line. 

Also in the D$_1$ line, the presence of atomic polarization in the lower 
$F$-levels produces an increase of the amplitude of the $Q/I$ signal. 
Although both the positive and negative peaks are increased, the variation is 
sensibly larger in the negative one.
Interestingly, the small negative structure in the blue wing is not 
affected by lower level polarization.
Contrary to the D$_2$ line, the impact of lower level polarization is almost 
negligible for $b=10$, while for $b=0.1$, the profile starts being very similar 
to the one calculated by \citet{Lan98}.

\begin{figure}
\centering
\includegraphics[width=0.9\textwidth]{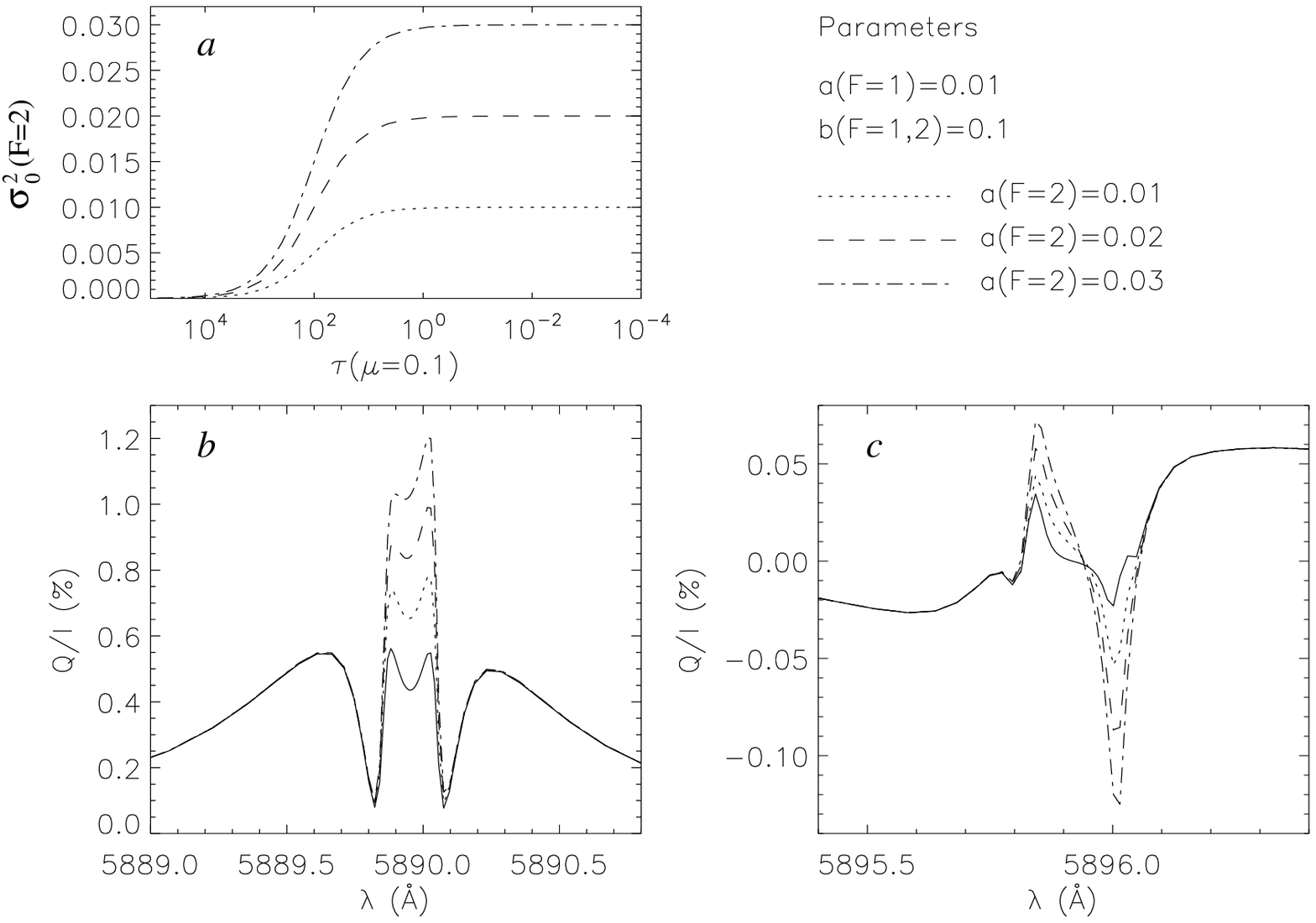}
\caption{{\it Panel a:} plot of $\sigma^2_0(F_{\ell}=2)$ as a function of 
optical depth (along the line-of-sight, at the line center frequency of D$_2$) 
in the FAL-X model, for three different values of the parameter 
$a(F_{\ell}=2)$. The values of the parameters $a$ and $b$ in the three 
cases considered are indicated in the figure. 
{\it Panel b:} fractional linear polarization profiles of the D$_2$ line, 
calculated for the three different parametrizations of lower level 
polarization shown in panel $a$.
The profile with the solid line corresponds to the case of no lower level 
polarization (it coincides with the dotted profile in panel $b$ of Fig.~1), 
and it is included for reference.
All of the profiles have been calculated in the FAL-X model, for $\mu=0.1$.
{\it Panel c:} same as panel $b$, but for the D$_1$ line.}
\label{Fig:llp_2}
\end{figure}

Figure~\ref{Fig:llp_2} shows the $Q/I$ profiles of the D$_1$ and D$_2$ lines 
calculated in the FAL-X model, for different values of the parameter 
$a(F_{\ell}=2)$, assuming $a(F_{\ell}=1)=0.01$ and 
$b(F_{\ell}=1)=b(F_{\ell}=2)=0.1$.
The amplitudes of the central peak of D$_2$, and of the positive and negative 
peaks of the $Q/I$ signal of D$_1$ increase proportionally to the atomic 
polarization of the lower level $F_{\ell}=2$.
As in the previous case, the variation of the signal amplitude in the D$_1$ 
line is larger in the negative peak than in the positive one.
Interestingly, the sub-structure of the central peak of D$_2$ becomes more 
asymmetric when increasingly different amounts of atomic polarization in the 
lower levels $F_{\ell}=1$ and $F_{\ell}=2$ are considered.

\subsection{Center-to-limb variation}
\label{Sect:CLV}
\begin{figure}[!ht]
\centering
\includegraphics[width=0.9\textwidth]{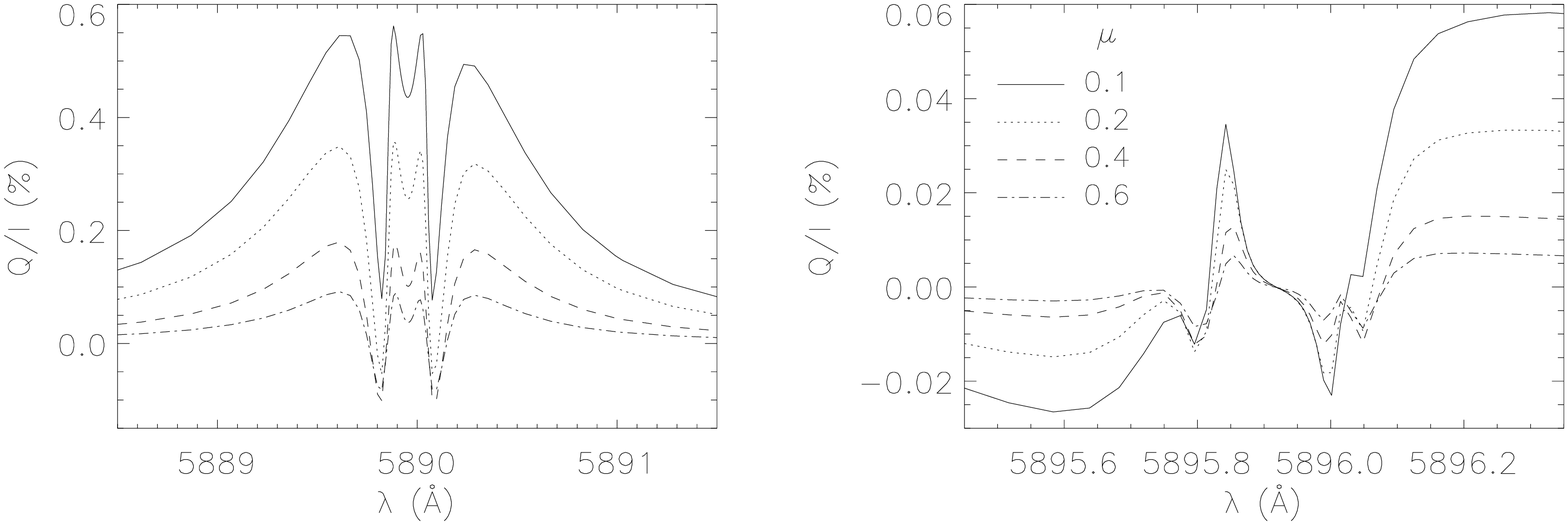}
\caption{{\it Left panel:} center-to-limb variation of the D$_2$ line $Q/I$ 
profile calculated in the FAL-X atmospheric model, in the absence of atomic 
polarization in the lower level.
The values of $\mu$ corresponding to the various profiles are indicated in the 
right panel.
{\it Right panel:} same as left panel, but for the D$_1$ line.}
\label{Fig:clv}
\end{figure}
Moving from the limb to disk center, the amplitude of the central peak and 
of the wing peaks of the D$_2$ line $Q/I$ profile gradually decreases (see left 
panel of Fig.~\ref{Fig:clv}).
The value of the two dips between the central peak and the wing peaks changes 
from positive to negative while going from $\mu=0.1$ to $\mu=0.2$. 
It reaches a (negative) minimum around $\mu=0.4$, and it finally starts 
decreasing (in absolute value), going to zero, for larger values of $\mu$.
The figure shows the $Q/I$ profiles up to $\mu=0.6$. For larger $\mu$-values
the whole profile goes gradually to zero without changing its shape and sign.

As far as the D$_1$ line is concerned, it can be observed that the amplitude of 
both the positive and negative peaks gradually decreases going from the limb to disk center (see right panel of Fig.~\ref{Fig:clv}).
The small negative dip in the blue wing of the line remains almost unchanged 
going from $\mu=0.1$ to $\mu=0.4$, and only for higher values of $\mu$ its 
amplitude starts decreasing.
Interestingly, in the red wing of the line, where, for $\mu=0.1$, a small and
almost flat positive feature is obtained, going to $\mu=0.2$ we find another 
small, but appreciable, negative dip. This dip does not change appreciably 
going from $\mu=0.2$ to $\mu=0.6$.
As for the case of the D$_2$ line, for values of $\mu$ larger than 0.6 (not 
shown in the figure), the whole $Q/I$ profile of the D$_1$ line goes gradually 
to zero without changing its shape and sign.

\section{Concluding comments}
The modeling of the linear polarization produced by resonance scattering in the 
solar atmosphere is a complex radiative transfer problem, especially when 
strong spectral lines resulting from HFS multiplets are considered. 
This is because there are, in general, several mechanisms and physical 
ingredients that need to be taken into account for explaining the observed 
spectral line polarization: frequency correlations between the incoming and 
outgoing photons along with the spectral structure of the incident radiation 
field, ground-level polarization, and quantum interference among FS and HFS 
levels. 
In this work, we have developed a theoretical and numerical approach suitable 
for solving the non-LTE radiative transfer problem for polarized radiation, 
taking the above-mentioned ingredients into account.

The theoretical approach is based on the density-matrix metalevel theory
proposed by \citet{Lan97}, according to which each atomic level is considered
as a continuous distribution of sublevels.
We consider a two-term atomic model with HFS, with prescribed atomic
polarization in the $F$-levels of the ground level, and we focus on the limit 
of coherent scattering in the atomic rest frame, taking into account the 
effects of Doppler redistribution in the observer's frame. 
Moreover, in addition to the radiative transitions we include excitations and 
de-excitations due to inelastic collisions with electrons, as explained in 
\citet{Bel15}. 
As far as elastic collisions with neutral hydrogen atoms are concerned, in this 
first step, we have neglected them, except for their line broadening effect 
(see footnote~\ref{FN:el_coll}).
With these assumptions, radiative transfer applications aimed at modeling the 
fractional scattering polarization observed in strong resonance lines are 
expected to be appropriate concerning the core of the lines. 
The numerical approach is a careful generalization of the methods explained in 
\citet{Bel14}.         

A detailed application to the D-lines of Na~{\sc i}, with emphasis on the 
enigmatic D$_1$ line, has allowed us to analyze the observable signatures of 
all the above-mentioned physical mechanisms. 
In agreement with \citet{Bel13b}, we conclude that the enigmatic linear 
polarization observed in the core of the sodium D$_1$ line may be 
explained by the effect that one gets when taking properly into account the 
detailed spectral structure of the incident solar D$_1$-line radiation
over the small frequency interval spanned by the HFS transitions. 
Interestingly, this key mechanism is capable of introducing significant 
scattering polarization in the core of the Na~{\sc i} D$_1$ line without the 
need for ground-level polarization. 

\acknowledgements
{\small IRSOL gratefully acknowledges financial support from the Swiss State
Secretariat for Education, Research and Innovation (SERI), Canton Ticino, the
city of Locarno, local municipalities, Aldo e Cele Dacc\`o foundation, and
the Swiss National Science Foundation through grant 200020-157103
(Astrophysical Spectropolarimetry).
Financial support by the Spanish Ministry of Economy and Competitiveness 
through projects AYA2010-18029 (Solar Magnetism and Astrophysical 
Spectropolarimetry) and AYA2014-60476-P is also gratefully acknowledged.}

\end{document}